\documentclass[11pt]{article}
\usepackage{graphicx}    
\usepackage{caption}
\usepackage{tabularx}
\usepackage{amsfonts}
\usepackage{bm}
\usepackage{amsmath}
\usepackage{lscape}
\usepackage{floatrow}
\usepackage{color}
\newcommand {\be}[1]{\begin{equation}\label{#1}}
	\newcommand {\ee}{\end{equation}}
\newcommand {\bea}{\begin{eqnarray}}
	\newcommand {\eea}{\end{eqnarray}}

\newcommand{\qed}{\hfill \rule{7pt}{7pt}}

\newtheorem{theorem}{Theorem}
\newtheorem{lemma}{Lemma}
\newtheorem{Defi}{Definition}

\newtheorem{prop}{Proposition}
\newtheorem{coro}{Corollary}

\usepackage[a4paper]{geometry}
\newfloatcommand{capbtabbox}{table}[][\FBwidth]
\usepackage{amssymb}
\usepackage{blindtext}
\usepackage{float}
\usepackage{array}
\usepackage{placeins}
\newcolumntype{L}{>{\centering\arraybackslash}m{3cm}}
    
\usepackage{setspace}

\title{Sequential unanimity voting rules for binary social choice}
\author{Stergios Athanasoglou\thanks{Department of Economics, University of Milan - Bicocca, \texttt{stergios.athanasoglou@unimib.it}.} \; \; Somouaoga Bonkoungou\thanks{Department of Economics, University of Lausanne, \texttt{bkgsom@gmail.com}. The authors would like to thank Carlos Alos-Ferrer, Herve Moulin, and seminar participants at the University of Milan-Bicocca and the Lausanne Theory meeting for useful comments. Bonkoungou acknowledges financial support from the Swiss National Science Foundation, project number \textrm{100018\textunderscore207722}.}}

\begin{document}
	\date{April 2024}
	\maketitle
	
	\begin{abstract}
		We consider a group of voters that needs to decide between two candidates. We propose a novel family of neutral and strategy-proof rules, which we call sequential unanimity rules. By demonstrating its formal equivalence to the M-winning coalition rules of Moulin \cite{m83}, we obtain a novel characterization of neutral and strategy-proof rules for binary social choice. 
		We establish our results by developing algorithms that transform a given M-winning coalition rule into an equivalent sequential unanimity rule and vice versa. Since M-winning coalition rules are identical to strong simple games, the analysis is relevant to this strand of the game-theoretic literature as well. Finally, our approach can be extended to accommodate the full preference domain in which voters may be indifferent between candidates.
		
	\end{abstract}

	\section{Introduction}
	$ $ \par
	We study voting rules for settings featuring two alternatives, or candidates. Despite its simplicity, this framework has many practical applications. In politics, elections held in two-party systems (parliamentary or presidential), as well as referendums, determine the outcome of a binary choice. In criminal legal proceedings, juries must render a verdict on a defendant's guilt. The determination of monetary policy, in which a group of central bankers needs to decide between two or three alternatives (raise, lower, or keep steady interest rates), provides yet another example of this kind of discrete decision-making process~\cite{rr10}.
	
	A voting rule is a function that picks a winning candidate for each profile of individual preferences. Correspondingly, the selection of a suitable voting rule is informed by the fairness, efficiency, and non-manipulability criteria that it satisfies.  A rule is {\em strategy-proof} if it is not possible for any voter to profitably misrepresent his preferences. A rule is {\em neutral} if it treats all candidates equally, and thus does not systematically favor one over another. Neutrality is a salient fairness principle in contexts where candidates represent entities that are harmed by discriminatory practices (political candidates, job applicants, etc.). A rule is {\em anonymous} if it treats all voters equally, rendering all aspects of their identity irrelevant to the voting process. 
	
	In a celebrated contribution, May~\cite{m52} showed that majority rule is characterized by anonymity, neutrality, and positive responsiveness, a straightforward monotonicity property that reduces to strategy-proofness when preferences are strict. This result, known as May's theorem, provided solid axiomatic foundations for majoritarian rules and had a major impact on the subsequent literature. However, despite the theoretical and intuitive appeal of majority rule, there are a number of settings in which voters are granted differential power. First, few actual political systems are consistent with pure majority rule. Any system that disaggregates the electorate into districts, whose elected representatives form a centralized body such as an assembly or parliament, will involve the violation of anonymity to some degree (Bartholdi et al.~\cite{b21}). This is also true of presidential systems of government that decompose the electorate into blocks and then aggregate the block-level results\footnote{The United States Electoral College, which grants outsize influence to the voters of a handful of swing states, is a prominent case in point.} Second, in many contexts, seniority or expertise is a valuable dimension in the decision process that justifies differences in voting power among agents. 
	
	Motivated by the above observations, we relax anonymity and study strategy-proof and neutral rules. To begin, we focus on the strict-preference case and consider M-winning coalition rules. First introduced by Moulin~\cite{m83}, this important family of rules represents the springboard for our analysis. A subset of voters $C$ is a {\em winning coalition} of a rule if, whenever all voters in $C$ prefer the same candidate, and all other voters prefer the other candidate, then the rule picks $C$'s preference.  An {\em M-winning coalition set} is a collection of winning coalitions that satisfies two properties: {\em minimality} and {\em Moulin's property}.\footnote{A collection of winning coalitions is {\em minimal} if no winning coalition within it is a proper subset of another. A collection of winning coalitions satisfies {\em Moulin's property} if, for any subset of voters $C$, we have that $C$ has a non-empty intersection with all elements of the collection if and only if it is a superset of an element of the collection. See Definition~\ref{def:wcs} for a clear formulation.} {\em M-winning coalition rules}, of which majority rule is a special case, are parameterized by an M-winning coalition set $\mathcal C$. For each profile of preferences, there is at least one element of $\mathcal C$ whose members are unanimous in their choice of preferred candidate and there can be no two elements of $\mathcal C$ who have different unanimous preferences. The M-winning coalition rule associated with $\mathcal C$ selects for each profile of preferences exactly that unique unanimous outcome. Minimality and Moulin's property ensure that the rule is well-defined, neutral and strategy-proof. It is important to note that M-winning coalition rules are identical to {\em strong simple games}, a topic in cooperative game theory that has received considerable attention in the literature (Taylor and Zwicker~\cite{tz99}).
	
	
	\paragraph{Contributions.} We propose a new family of rules, {\em sequential unanimity rules}, that are equivalent to M-winning coalition rules but arguably simpler to describe and implement. Sequential-unanimity rules are parameterized by a sequence $\textrm S=(S_1,...,S_K)$ of non-recurring voter subsets whose last element is a singleton. Briefly, here is how they work. Given a profile of preferences, a sequential unanimity rule examines the first subset in its associated sequence $\textrm S$. If the preferences of the voters composing it are unanimous, the rule stops and outputs this preferred candidate. Otherwise, it considers the next element in the sequence $\textrm S$ and repeats the procedure until it encounters a group of voters who all agree on their preferred candidate. At that point, the process terminates and the rule outputs this candidate. 
	
	Sequential-unanimity rules have several advantages compared to M-winning coalition rules. When the number of candidates exceeds a modest threshold, the definition of M-winning coalition rules becomes more complex. This is because the regularity requirements of minimality and Moulin's property can be hard to verify as the problem size grows. Moreover, even if regularity requirements were met, implementing M-winning coalition rules entails two demanding tasks, at least in principle. First, we would need to elicit each voter's preference. In cases where the determination of a preference entails significant effort (e.g. the assessment of job market candidates in academia), this elicitation imposes a cognitive burden and an opportunity cost on the voters. Second, we would need to search for an element of the M-winning coalition set featuring unanimous agreement. While this is guaranteed to exist, if the cardinality of the M-winning coalition set is large, and its structure is complex, the process may be daunting. By contrast, sequential unanimity rules do not require any regularity conditions and can be straightforwardly implemented, even when the number of candidates is large.
	
	Despite their inherent differences, we are able to establish an intimate connection between M-winning coalition rules and sequential unanimity rules. We do so via a constructive approach. First, we develop an algorithm (``Algorithm 1") that transforms a given M-winning coalition rule into an equivalent sequential unanimity rule. Algorithm 1 is a nontrivial procedure, in that it never results in a simple enumeration of all the elements of an M-winning coalition set plus a final backstop. Combined with Moulin's characterization of M-winning coalition rules~\cite{m83}, this result delivers a novel characterization of strategy-proof and neutral rules for binary social choice. Subsequently, we propose a different algorithm (``Algorithm 2") that transforms a given sequential unanimity rule into an equivalent M-winning coalition rule. We thus have a complete equivalence result between these two families of rules. In addition, upon slightly modifying the two algorithms, the equivalence carries over to the full preference domain, which allows for voters to be indifferent between candidates.\footnote{We stress that our results do not imply a characterization of these rules in the full domain case. Indeed, we are not aware of any characterization of neutral and strategy-proof rules when voters can be indifferent between the two candidates.}
	
	The above results are potentially relevant to the theory of strong simple games, since they suggest a systematic sequential procedure for determining their outcome. This is particularly important for strong simple games that are {\em not weighted}. In such games, it is impossible to assign weights to the voters and define winning coalitions as consisting of those subgroups whose aggregate weight exceeds some threshold~\cite{tz99}. While weighted schemes are considerably more intuitive and widespread, non-weighted systems are not a mere theoretical curiosity. Several actual voting procedures, including the United States Federal System and the European Union Council, correspond to non-weighted strong simple games (see Examples 1.4-1.5-2.4 in Freixas and Molinero~\cite{fm09}). In principle, our Algorithm 1 provides a promising tool for simplifying such games and computing their outcome in a faster way. We demonstrate its application in one such example in Section 3.1. 
	
	\paragraph{A concrete example.} The following example is useful in illustrating the paper's framework and basic results. Suppose we have a panel of seven professors who need to decide between two job market candidates, Ann and Bob. Professors 1 and 2 are full professors, 3 and 4 are associate professors, and 5,6 and 7 are assistant professors. Reflecting differences in seniority and status, the opinion of a full professor carries twice the weight of that of an associate, and the opinion of an associate carries twice the weight of an assistant. The decision process is such that the candidate who enjoys the support of at least two full professors, or their weighted equivalent, will be selected.

	
	While not immediately obvious, the above rule is an instance of a M-winning coalition rule having M-winning coalition set $\mathcal C=\{\{1,2\}, \{1,3,4\}, \{2,3,4\}, \{1,3,5,6\}, \{1,3,5,7\}, \{1,3,6,7\},\\ \{1,4,5,6\}, \{1,4,5,7\}, \{1,4,6,7\}, \{2,3,5,6\}, \{2,3,5,7\},  \{2,3,6,7\},\{2,4,5,6\}, \{2,4,5,7\},\\ \{2,4,6,7\}\}$.\footnote{Equivalently, this setting can be interpreted as a weighted strong simple game with weight 4 for voters 1 and 2, weight 2 for voters 3 and 4, and weight 1 for voters 5, 6, and 7. The aggregate weight threshold defining a winning coalition is 8.} For instance, if professors 1 and 2 both prefer Ann to Bob, Ann is selected. Conversely, if professors 1,3,5, and 6 prefer Bob to Ann, then Bob receives support from 1 full professor, 1 associate professor, and two assistant professors. This adds up to the weighted equivalent of two full professors and thus Bob is selected.

	
	One way to implement this rule would be to ask all the professors to assess the candidates, search for an element of $\mathcal C$ whose members have unanimous preferences, and select the corresponding candidate. However, candidate assessment is a time-consuming and cognitively demanding task. Asking for all professors to weigh in is not optimal: ideally, we would like to ask for a faculty member's opinion only if it is needed. With this in mind, an alternative, and arguably preferable, approach is to proceed sequentially in the following manner: first, solicit the opinion of professors 1 and 2. If they are unanimous, then the winning candidate is determined and we can stop there. If not, consult professors 3 and 4 and if {\em they} are unanimous, then once again the winner is determined (for she or he will have the support of one full and two associate professors, which add up to an equivalent of two full professors). Conversely, if professors 3 and 4 are not unanimous, then proceed with professors 5 and 6 and repeat: if they are unanimous, then once again the winner is determined. Otherwise, we move on and terminate the procedure by simply picking Professor 7's preferred candidate.
	
	It is not difficult to see that the above sequential unanimity rule (which we can obtain via our Algorithm 1 applied to $\mathcal C$) will always produce the same outcome as the M-winning coalition rule associated with $\mathcal C$. Moreover, it has two distinct advantages compared to it: (i) it is likely to solicit the input of a restricted set of professors and (ii) it does away with the need to search among the 15-element coalition set $\mathcal C$, instead disciplining the process to a sequential search of at most four voter subsets.\footnote{The M-winning coalition set in this example has a nice structure due to the clearly defined hierarchy of voters and weights. It is tempting to conclude that M-winning coalition sets should always be this orderly. But this is not the case, even when the underlying strong simple game is weighted.}
	
	\paragraph{Related work.} Our work is primarily relevant to the literature on binary social choice. As mentioned earlier, the canonical contribution in this area is May's Theorem~\cite{m52}, which provides a firm axiomatic foundation for majority rule. Subsequently, drawing on work by Fishburn and Gehrlein \cite{fg77}, Moulin~\cite{m83}  (page 64) showed that M-winning coalition rules uniquely satisfy strategy-proofness and neutrality. Along related lines, Larsson and Svensson~\cite{ls06} axiomatized a different but similar family of rules, known as {\em voting by committee}, with strategy-proofness and ontoness. 
	
	
	There are a number of papers that examine the implications of anonymity for strategy-proof social choice in the two-candidate context. Lahiri and Pramarik~\cite{lp20} generalized the characterization of Larsson and Svensson in a model with indifferences. In a series of recent papers, Basile et al.~\cite{brr21, brr22a, brr22b, brr22c} provided alternative characterizations of strategy-proof anonymous rules that are simpler than Lahiri and Pramarik~\cite{lp20} and admit explicit functional forms. To address the definitional complexity of the rules of Lahiri and Pramarik~\cite{lp20}, Basile et al.~\cite{brr22a} offered an alternative, explicit characterization that is easier to work with. In a companion paper, Basile et al.~\cite{brr21} extended this analysis to the case in which the voters must decide between two candidates but can indicate preferences over a wider set. In Basile et al.~\cite{brr22c} they developed a unified framework that leads to a representation of various classes of strategy-proof rules, including anonymous ones.
	
	Anonymity, while intuitive, is rarely satisfied in political systems based on some form of representative democracy. In such settings weaker standards of inter-voter fairness apply. Motivated by this observation, Bartholdi et al.~\cite{b21} proposed $k$-equity, a parametric relaxation of anonymity. This weaker requirement leads to the identification of a rich family of equitable rules that satisfy neutrality and positive responsiveness. Using insights from group theory, Bartholdi et al.~\cite{b21} were able to establish lower and upper bounds of magnitude $O(\sqrt{n})$ on the size of the winning coalitions defining such rules. Their work implies that consensus between a relatively small number of voters can be sufficient to determine the outcome of voting systems that meet reasonable standards of fairness. Working within a similar framework as Bartholdi et al., Kivinen~\cite{k23} studied the equity-manipulability tradeoff inherent in the design of voting rules. He introduced a slightly different version of equity, pairwise equity, and showed that for rules satisfying it anonymity is equivalent to various measures of group strategy-proofness.
	
	As mentioned before, our analysis is also relevant to the literature on strong simple games \cite{tz99}. Necessary and sufficient conditions for a strong simple game to be weighted have been the object of consistent study in the game-theoretic~\cite{tz93,tz96,hz14} and discrete-mathematics~\cite{tz92,tz95,fm09} literature. It is not immediately clear how these insights affect the analysis in our paper, in particular the impact they might have on the structure and implementation of Algorithms 1 and 2. Getting a clearer picture of the connection between our analysis and the theory of strong simple games constitutes an interesting avenue of future research.
	
	

	\section{Model}	
	Let $N$ be a finite set of $n$ voters and $A=\{a,b\}$ a set of two candidates. Voter $i$'s preference relation over candidates is denoted by $R_i$ so that $R_i=x$ for some $x\in A$ means that voter $i$ prefers candidate $x$ to candidate $y\neq x$. A {\bf profile} $R_N$ is an n-tuple of preferences and the space of profiles is denoted by $\mathcal{R}^N$. A {\bf rule} is a function $f: \mathcal{R}^N\rightarrow A$ that assigns a candidate to each profile.

	\begin{Defi} \label{def:stp}
	A rule $f$ is {\bf strategy-proof} if for all profiles $R_N \in \mathcal R^N$, each voter $i \in N$ and each $R'_i\neq R_i$, $i$ does not prefer $f(R'_i,R_{-i})$ to $f(R_N)$.
	\end{Defi}

	Given a permutation $\pi: A\rightarrow A$ of $A$ and a profile $R_N$, define the profile $\pi R_N$ as $\pi R_i = \pi (R_i)$ for all $i \in N$. 
	
	\begin{Defi} \label{def:neut}
	A rule $f$  is {\bf neutral} if for each permutation $\pi$ of $A$, and all profiles $R_N \in \mathcal R^N$,
	\begin{equation*}
		f(\pi R_N)=\pi(f(R_N)).
	\end{equation*}
	\end{Defi}
	
	We now introduce the concept of a winning coalition of a rule.

	\begin{Defi}
		A subset $C \subset N$ of voters\footnote{For clarity, we specify that in our paper the symbol $\subset$ denotes ``subset of", while $\subsetneq$  denotes ``proper subset of".}  is a {\bf winning coalition} of rule $f$ if for all profiles $R_N\in \mathcal{R}^N$ satisfying $R_i=x$ for all $i \in C$ and $R_i \neq x$ for all $i \not \in C$, we have $f(R_N)=x$.  
	\end{Defi}
	
	The attentive reader will notice that the above definition of winning coalitions is consistent with Moulin's~\cite{m83} and slightly different from Bartholdi et al.~\cite{b21}.\footnote{They substitute the statement ``$R_i=x$ for all $i \in C$ and $R_i \neq x$ for all $i \not \in C$'' with ``$R_i=x$ for all $i\in C$''.} The two definitions coincide under the added assumption of strategy-proofness or positive responsiveness.

	\begin{Defi}
		\label{def:wcs}
		A non-empty set $\mathcal C$ of non-empty subsets of $N$ is an {\bf M-winning coalition set} if it satisfies the following two properties:
		\begin{itemize}
			\item [P1.] for all $C, C' \in \mathcal{C}$ we have $C \subset C' \Rightarrow C=C'$ [Minimality];
			\item[P2.] for all $C \subset N$, we have $\left [ C \supset C' \; \;  \textrm{for some} \; \; C' \in \mathcal C\right]  \Leftrightarrow \left[ C \cap C'' \neq \emptyset \; \; \textrm{for all} \; \; C'' \in \mathcal C\right]$ [Moulin's Property].
		\end{itemize}
	\end{Defi}
	
	The structure of M-winning coalition sets can range from highly ordered and symmetric to apparently quite random. For example,
	
	\begin{itemize}
		\item [1.] $\mathcal C^1= \{ C \subset \{1,2,...,9\}: \; \; |C|=5\}$ and
		\item [2.] $\mathcal C^2=\{ \{1,2,3\}, \{1,2,4\}, \{1,2,7\}, \{2,3,4\},\{1,3,5,6\},\{1,3,5,7\}, \{1,3,6,7\}, \{1,4, 5,6,8\} \\ \{2,3,5\},\{2,3,6\}, \{2,3,8\}, \{2,5,7\},\{2,6,7\},  \{2,4,5,6\}, \{3,4,7\} \}$
	\end{itemize}
	are both valid M-winning coalition sets.

	\begin{Defi}\label{def:wcr}
		Consider an M-winning coalition set $\mathcal C$. The {\bf M-winning coalition rule}  $\mathcal C: \mathcal R^N \mapsto A$, satisfies, for all $R_N \in \mathcal R^N$,
		$$\mathcal C(R_N)=x \; \Leftrightarrow \; \exists C \in \mathcal C \textrm{  s.t.  } R_i=x \textrm{   for all  } i \in C.$$
	\end{Defi}
	
	First introduced by Moulin \cite{m83}, M-winning coalition rules are well-defined. We briefly demonstrate why this is true. First, we claim that for any profile $R_N$, there exists $C\in \mathcal{C}$ and $x \in A$ such that $R_i=x$ for all $i \in C$. To see this, let $N^a=\{i \in N:\; R_i=a\}$ and $N^b=\{i \in N:\; R_i=b\}$. We consider two cases. First, suppose that $N^a$ has a non-empty intersection with every element of $\mathcal{C}$; that is, for all $C\in \mathcal{C}$, $N^a\cap C\neq \emptyset$. By property P2, there exists $C\in \mathcal{C}$ such that $N^a \supset C$, proving that $\mathcal C(R_N)=a$. Second, suppose that there exists $C\in \mathcal{C}$ such that $N^a\cap C=\emptyset$. Then $C\subset N^b$ and for all $i \in C$ we have $R_i=b$. This implies that $\mathcal C(R_N)=b$. Finally, we claim that there are no $C, C'\in \mathcal{C}$ such that $R_i=a$ for all $i \in C$ and $R_j=b$ for all $j \in C'$. This holds because property P2 implies $C \cap C''\neq \emptyset$ for all $C'' \in \mathcal C$ --including $C'$. In conclusion, for all profiles $R_N \in \mathcal R^N$, there exists $C\in \mathcal{C}$, and $x\in A$ such that $R_i=x$ for all $i \in C$ and there is no other $C'\in \mathcal{C}$ such that $R_j=y\neq x$ for all $j \in C'$.
	
	As mentioned in the introduction, M-winning coalition rules are characterized by strategy-proofness and neutrality.
	
	\begin{theorem}[Moulin \cite{m83}]
		A rule is strategy-proof and neutral if and only if it is an M-winning coalition rule.\label{th:moulin}
	\end{theorem}

	\paragraph{Connection to strong simple games.} We end by noting that M-winning coalition rules are formally identical to the concept of {\em strong simple games} in cooperative game theory~\cite{tz99}.\footnote{We thus occasionally use the two terms interchangeably.} This allows us to draw connections to the simple game literature and leverage its theoretical insights. A strong simple game is defined as a pair $(N, \mathcal C)$, where $\mathcal C$ is an M-winning coalition set on $N$. A central issue in that literature is the property of weightedness.  We say that a strong simple game $(N,\mathcal C)$ is {\bf weighted} if there exists a vector of weights $\bm w \in \Re^{|N|}$, where $w_i$ is the weight of voter $i\in N$, and a threshold $q \in \Re$, such that $\mathcal C= \{C \subset N: \sum_{i\in C}w_i \geq q\}$.\footnote{Note that $\bm w$ and $q$ have to be carefully set, so as to avoid having two subsets $C, C' \in \mathcal C$ such that $C \cap C' = \emptyset$.} A strong simple game that does not admit such a representation is {\bf non-weighted}. 
	
	The job-market candidate example of the Introduction is an example of a weighted strong simple game with weight vector $\bm w=(4,4,2,2,1,1,1)$ and threshold $q=8$. In addition, the $M$-winning coalition set $\mathcal C^1$, right before Definition~\ref{def:wcr}, is yet another example of such a game with $\bm w=(1,1,1,1,1,1,1,1,1)$ and $q=5$.
	
	Weighted strong simple games constitute the canonical example of neutral and strategy-proof rules for binary social choice. But, not all strong simple games are weighted and a considerable amount of research has been dedicated to ascertaining the exact conditions that render a game weighted or not~\cite{tz92,hz14,fm09}. A prominent necessary and sufficient condition for a strong simple game to be weighted is the property of {\em trade robustness}, proposed by Taylor and Zwicker~\cite{tz92}. Without giving a formal definition, the essence of trade robustness is that it is not possible to transform a set of winning coalitions into a set of losing coalitions via a sequence of inter-coalitional voter trades. The strong simple game $\mathcal C^2$ mentioned earlier is not weighted because it fails exactly this criterion. To see this, consider the winning coalitions $C_1=\{2,3,6\}$ and $C_2=\{3,4,7\}$, which both belong to $\mathcal C^2$. Suppose voters 6 and 7 flip coalitions, resulting in the updated subsets $C_1'=\{2,3,7\}$ and $C_2'=\{3,4,6\}$. Since $C_1' \not \in \mathcal C^2$ and $C_2' \not \in \mathcal C^2$, the property of trade robustness is violated. Hence, the strong simple game $(N,\mathcal C^2)$ is non-weighted.

	\section{Sequential Unanimity Rules}
	
	The next family of rules we will focus on is parameterized by a sequence $\textrm S=(S_1,\hdots, S_K)$ of non-empty, non-recurring subsets of $N$ such that $\lvert S_K\lvert =1$. We suppose, moreover, that the voter in the singleton $S_K$ does not belong to any other set in the sequence. The {\bf sequential unanimity rule} $\mathcal S$ considers each set in the sequence $\textrm S$ in increasing order (from $S_1$ until $S_K$) and examines the preferences of the voters belonging to it. As soon as a set is encountered in which voter preferences are unanimous, the procedure stops and the outcome of rule $\mathcal S$ is exactly that candidate for which there is unanimous support. The following definition formalizes this idea.
	
	\begin{Defi}
		\label{def:sur}
		Consider a sequence $S=(S_1,...,S_K)$ of non-empty, non-recurring subsets of $N$ satisfying $|S_K|$=1. The {\bf sequential unanimity rule} $\mathcal S: \mathcal R^N \mapsto A$ is defined as follows: for all $R_N \in \mathcal R^N$,
		\begin{eqnarray*}
			\mathcal S(R_N)=x \; \;  \Leftrightarrow  & & \bigg [ \exists \; k\in \{1,...,K\} \textrm{  s.t.  } R_i=x \textrm{   for all  } i \in S_k  \\
			& & \textrm{  and  for all $l < k$ there exist $i,j \in S_l$ s.t. $R_i \neq R_j$.} \bigg ]  \end{eqnarray*}
	\end{Defi}

	\begin{prop}
		Sequential unanimity rules are neutral and strategy-proof.\label{prop:seq}
	\end{prop}
	
	\noindent{\bf Proof.} It is easy to show that sequential unanimity rules are neutral. To establish their strategy-proofness, consider a sequential unanimity rule $\mathcal S$ with associated sequence S=$(S_1,...,S_K)$. Suppose that $\mathcal{S}(R_N)=x \in A$ and that subset $S_l$ is the first element of sequence $\textrm S$ to feature unanimous agreement. Now consider a voter $i\in N$ and suppose that $R_i\neq x$. This implies that $i\not \in S_l$. Suppose this voter misreports his preferences, stating $R_i'=x$. We distinguish between two cases. If $i\not \in S_{k}$ for all $k< l$, then voter $i$'s preferences are not consulted before the rule reaches an outcome, implying $\mathcal{S}(R'_i,R_{-i})=\mathcal{S}(R_N)=x$. Conversely, let $\mathcal K= \{k <l: \; i \in S_k\}.$ If there exists $k \in \mathcal K$ such that $R_j=x$ for all $j \in S_k$ such that $j\neq i$, then $\mathcal S(R_i',R_{-i})=x$. If not, then $S_l$ will once again be the first element of sequence $\textrm S$ to feature unanimous agreement, implying $\mathcal S(R_i',R_{-i})=x$. We conclude that $\mathcal S$ is strategy-proof. \qed

	\subsection{From M-winning coalition rules to sequential unanimity rules}
	
	In this section, we show how an M-winning coalition rule can be transformed into a family of equivalent sequential unanimity rules. Our approach is constructive as we develop an algorithm that takes as input an M-winning coalition set and produces a sequence of voter subsets. We refer to this algorithm as {\bf Algorithm 1}. A corollary of this algorithm is a characterization of the sequential unanimity rules produced via Algorithm 1 with neutrality and strategy-proofness.

	\subsubsection{General description of Algorithm 1}
	
	At a high level, Algorithm 1 works in the following way. Given an M-winning coalition set $\mathcal C$, it begins by choosing an arbitrary voter $i \in N$, which it designates as the ``backstop'' voter, and removes all elements of $\mathcal C$ that contain it. At each subsequent iteration, the algorithm further removes elements of $\mathcal C$. It does so by selecting a subset of voters $N'$ that satisfies three specific criteria: 
	\begin{itemize}
		\item[(i)] it has cardinality at least 2, 
		\item [(ii)] it is a proper subset of some remaining element of $\mathcal C$, and 
		\item [(iii)] it has a nonempty intersection with all the previously discarded elements of $\mathcal C$. 
	\end{itemize}
	Note that there may be more than subset $N' \subset N$ satisfying the above. In that case, the algorithm picks one arbitrarily, while respecting the following {\bf proviso}: if $N'$ satisfies (i)-(ii)-(iii) and there exists another $N'' $ doing the same such that $N' \subsetneq N''$, then subset $N''$ cannot be selected. This condition is imposed in order to avoid producing a sequence $\textrm S=(S_1,...,S_K)$ such that $S_l \subsetneq S_k$ for some $l<k$. A sequence of this kind would be inefficient, as subset $S_k$ is superfluous to the application of the rule. This is because the sequential unanimity rule associated with S can never reach an outcome immediately upon consulting the preferences of voters in $S_k$: (i) if voters in $S_l$ are not unanimous, then also voters in $S_k$ are not unanimous and (ii) if voters in $S_l$ are unanimous, then the rule terminates before consulting subset $S_k$.\footnote{Note that, unless the proviso is imposed, Algorithm 1 might produce a sequence having this inefficient feature. See section 3.1.4.}
	
	When we can no longer identify a subset of $N$ satisfying criteria (i)-(ii)-(iii), Algorithm 1 terminates. The output sequence S is constructed in the following way. In last place, we put the backstop voter. In the second to last place, we put subset $N_2$. Right before $N_2$ we put $N_3$, and so on, until the last such subset identified via the satisfaction of criteria (i)-(ii)-(iii). At that point, if the M-winning coalition set  $\mathcal C$ has not been completely exhausted, then its remaining elements are placed, in any order, at the beginning of the sequence $\textrm S$. 
	
	What should be clear is that when applying Algorithm 1 different choices of backstop voter and/or subsequent voter subsets will yield different sequences. Indeed, we can associate an entire family of sequences $S^{\mathcal C}$ to each M-winning coalition set $\mathcal C$.

	\begin{figure}[H]
		\centering
		\includegraphics[height=.2\textheight, width=.75\textwidth]{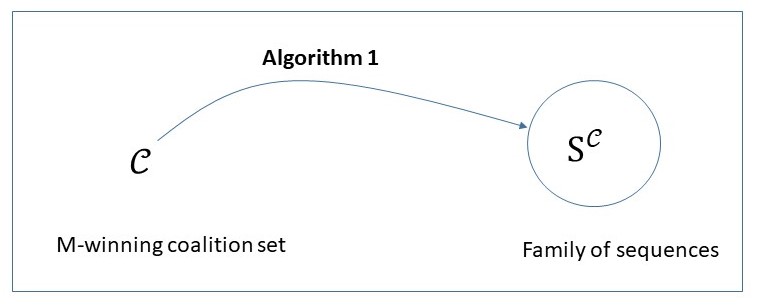}
		\caption{Different applications of Algorithm 1 (i.e. different choices of backstop and $N_k$ subsets during the course of the algorithm) associate to each M-winning coalition set $\mathcal C$ a family of sequences $S^{\mathcal C}$.}
		\label{fig:Alg1}
	\end{figure}
	
	We will show that the M-winning coalition rule $\mathcal C$ is equivalent to any sequential unanimity rule $\mathcal S$ having sequence $\textrm S \in S^{\mathcal C}$. Before doing so, and to generate some intuition for Algorithm 1, we illustrate it on the academic job market example discussed in the introduction.

	\subsubsection{An illustrative example}
	
	We have $N=\{1,2\hdots,7\}$ and the M-winning coalition set depicted in Figure \ref{fig:Alg1_1} below.
	
	\begin{figure}[H]
		\centering
		\includegraphics[height=.25\textheight, width=.4\textwidth]{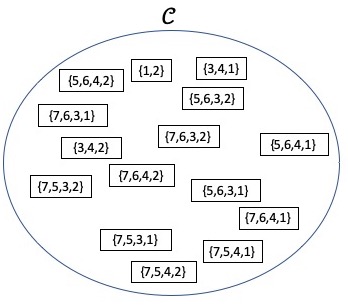}
		\caption{{\small M-winning coalition set $\mathcal C$.}}
		\label{fig:Alg1_1}
	\end{figure}
	
\begin{figure}[H]
		\centering
		\includegraphics[height=.22\textheight, width=.75\textwidth]{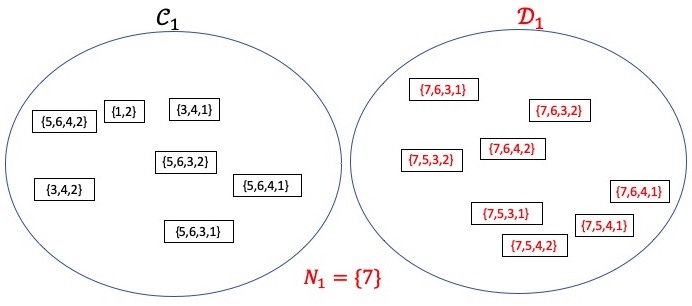}
		\caption{{\small Iteration 1: Set backstop voter $i=7$ and remove all elements of $\mathcal C$ containing it. The remaining M-winning coalition set is $\mathcal C_1$ and the discarded M-winning coalition set is $\mathcal D_1$.}}
		\label{fig:Alg1_3}
	\end{figure}

	We begin applying Algorithm 1 by selecting the backstop voter. While we could pick any voter in $N$ for this purpose, we want to be consistent with the introductory discussion and so we pick $N_1={7}$. Subsequently, we update the M-winning coalition set by removing all its elements containing voter 7. For clarity, we refer to the {\em remaining} M-winning coalition set as $\mathcal C_1$ and the set of {\em discarded} winning coalitions of $\mathcal C$ as $\mathcal D_1$. We illustrate in Figure \ref{fig:Alg1_3}.

\begin{figure}[H]
		\centering
		\includegraphics[height=.25\textheight, width=.75\textwidth]{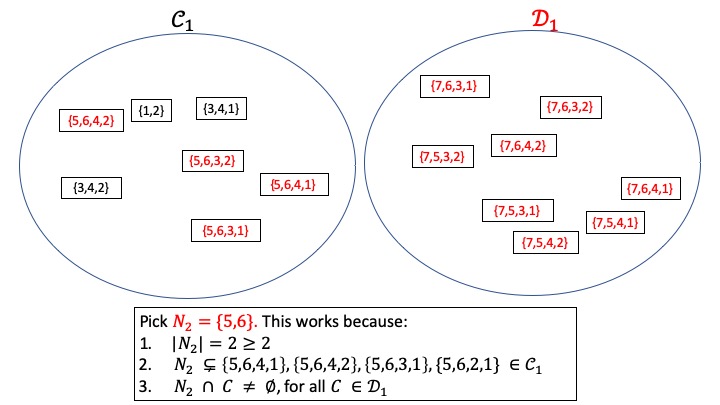}
		\caption{{\small Iteration 2: Selecting subset $N_2$.}}
		\label{fig:Alg1_4}
	\end{figure}

\begin{figure}[H]
		\centering
		\includegraphics[height=.22\textheight, width=.75\textwidth]{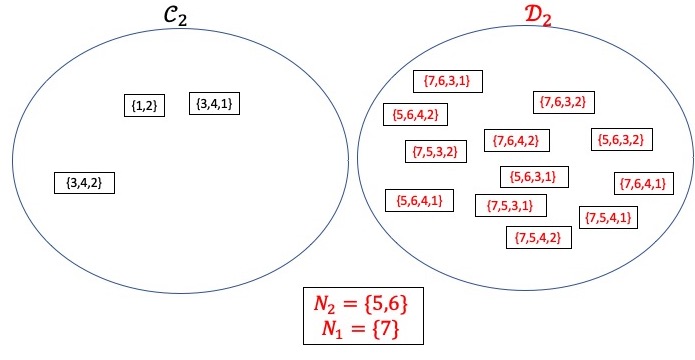}
		\caption{{\small Iteration 2: Updating sets $\mathcal C_1$ and $\mathcal D_1$ to $\mathcal C_2$ and $\mathcal D_2$.}}
		\label{fig:Alg1_5}
	\end{figure}

	We continue by searching for a subset of voters $N_2$ that satisfies the three criteria laid out in the previous subsection. There are various choices we could make at this point. Once again, for consistency with the introduction, we choose $N_2=\{5,6\}$. Figure \ref{fig:Alg1_4} highlights the elements of $\mathcal C_1$ that are proper supersets of $N_2$ and demonstrates why the subset $\{5,6\}$ is a is a valid choice.  Figure \ref{fig:Alg1_5} displays the corresponding updates of the remaining M-winning coalition set $\mathcal C_1$ and the discarded M-winning coalition set $\mathcal D_1$, to $\mathcal C_2$ and $\mathcal D_2$ respectively. To be precise, $\mathcal C_2 = \mathcal C_1 \setminus \{ \{5,6,3,1\}, \{5,6,3,2\}, \{5,6,4,1\}, \{5,6,4,2\}\}$ and $\mathcal D_2 = \mathcal D_1 \cup \{  \{5,6,3,1\}, \{5,6,3,2\}, \{5,6,4,1\}, \{5,6,4,2\}\}$. 
	
\begin{figure}[H]
		\centering
		\includegraphics[height=.25\textheight, width=.75\textwidth]{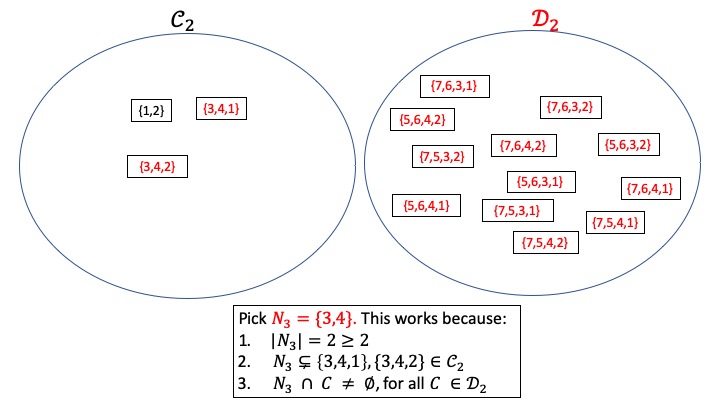}
		\caption{{\small  Iteration 3: Selecting subset $N_3$.}}
		\label{fig:Alg1_6}
	\end{figure}

\begin{figure}[H]
		\centering
		\includegraphics[height=.22\textheight, width=.75\textwidth]{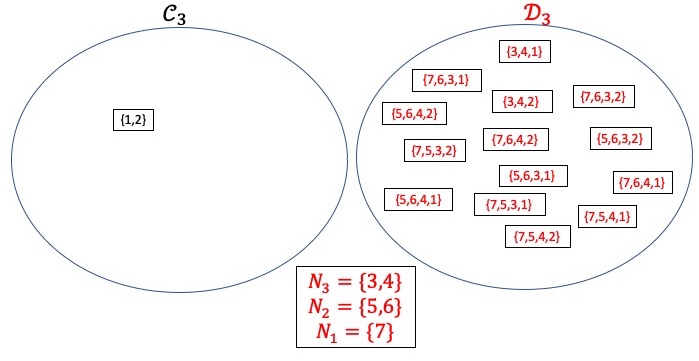}
		\caption{{\small Iteration 3: Updating sets $\mathcal C_2$ and $\mathcal D_2$ to $\mathcal C_3$ and $\mathcal D_3$.}}
		\label{fig:Alg1_7}
	\end{figure}

	We continue by searching for a subset of voters $N_3$ that satisfies the three criteria laid out in the previous subsection. For similar reasons as before, we choose $N_3=\{3,4\}$. Figure \ref{fig:Alg1_6} highlights the elements of $\mathcal C_2$ that are proper supersets of $N_3$ and demonstrates why the subset $\{3,4\}$ is a is a valid choice.  Figure \ref{fig:Alg1_7} displays the corresponding updates of the remaining M-winning coalition set $\mathcal C_2$ and the discarded M-winning coalition set $\mathcal D_2$, to $\mathcal C_3$ and $\mathcal D_3$ respectively. 
	
	At this point, there is a single element left in the remaining M-winning coalition set. Moreover, there is no way to satisfy all three criteria: in particular, criteria (i) and (ii) are mutually exclusive. Thus, Algorithm 1 terminates. The output sequence that it produces is the following:$$\textrm S^1=(\{1,2\}, N_3, N_2, N_1)= ( \{1,2\}, \{3,4\}, \{5,6\}, \{7\}).$$
	
	Before moving on to the next subsection, we briefly comment on how different ways of applying Algorithm 1 can lead to very different output sequences. To this end, keeping the same coalition set $\mathcal C$ of Figure \ref{fig:Alg1}, we re-apply Algorithm 1 twice. For additional details on how the resulting sequences $\textrm S^2, \textrm S^3$ are obtained, please consult the paper's Appendix A1.
	
	First, we begin by selecting voter 4 as the backstop, so that $N_1=\{4\}$. Subsequently, we set  $N_2=\{3,5,6\}, N_3=  \{3,6,7\}, N_4= \{3,5,7\}$. At that point, only subset $\{1,2\}$ remains and the algorithm terminates with the following output sequence:$$\textrm S^2=(\{1,2\}, N_4, N_3, N_2, N_1)= ( \{1,2\}, \{3,5,7\}, \{3,6,7\}, \{3,5,6\},\{4\}).$$To conclude, we demonstrate that the algorithm may occasionally terminate very quickly and with multiple remaining subsets. To wit, we select voter 2 as the backstop, so that $N_1=\{2\}$. After this point, there is no subset that will satisfy the three criteria laid out in the algorithm description. Seven subsets remain, namely all of the elements of $\mathcal C$ containing voter 1 except subset $\{1,2\}$. These can be ordered in an arbitrary fashion and placed in the first seven spots in the output sequence. Choosing one such ordering, we obtain the following sequence:$$\textrm S^3=(\{7,5,3,1\}, \{5,6,3,1\}, \{7,5,4,1\}, \{5,6,4,1\},\{7,6,3,1\},\{3,4,1\}, \{7,6,4,1\}, \{2\} ).$$

	\subsubsection{Algorithm 1 and its properties}
	
	We now give a formal description of Algorithm 1 and prove its properties.
	
	\begin{center}
		\fbox{\begin{minipage}{\linewidth}
				\noindent {\bf Algorithm 1$\big |$}\; \; \; \;  {\bf Input}: M-winning coalition set $\mathcal C$. \; \; {\bf Output}: Sequence $\textrm S$
				\hrule
				\begin{itemize}
					\item [1.] Pick an arbitrary $i \in N$ and define $N_1\equiv i$. Let$$\mathcal D_1= \{ C \in \mathcal C: \; N_1 \subset C\}, \; \; \; \mathcal C_1 = \mathcal C \setminus \mathcal D_1.$$
					\item [2.] For $k=2,3,...$, select $N_k \subset N$ such that the following three conditions hold:
					\begin{itemize}
						\item [(i)]  $|N_k| \geq 2$.
						\item [(ii)]  $N_k \subsetneq C$ for some $C \in \mathcal C_{k-1}$
						\item [(iii)] $N_k \cap C \neq \emptyset$ for all $C \in \mathcal D_{k-1}$. 
					\end{itemize}
					If multiple subsets satisfy the above properties, choose among them arbitrarily with the following {\bf proviso}: do {\bf not} pick a subset $N'$ if it is a proper superset of another subset that satisfies criteria (i)-(ii)-(iii). Define\begin{eqnarray*} \mathcal D_k & =& \mathcal D_{k-1} \cup \{C \in \mathcal C_{k-1}: \; N_k \subsetneq C\}, \\
						\mathcal C_k & = & \mathcal C_{k-1} \setminus \{C \in \mathcal C_{k-1}: \; N_k \subsetneq C\}.\end{eqnarray*}
					\item [3.] If no such $N_k \subset N$ exists, STOP. Let $k^* \equiv k$ and $\mathcal C^*\equiv \mathcal C_{k^{*}-1}$. Suppose $\mathcal C^*$ has $m$ elements and let $C_1^*,...,C_m^*$ be an arbitrary ordering of them.
					\item [4.] The output sequence is given by:$$\textrm S=\left( C_1^*,...,C_m^*, N_{k^*-1}, N_{k^*-2}, ..., N_2, N_1\right).$$
				\end{itemize}
		\end{minipage}}
	\end{center}
	
	\medskip

	\begin{theorem}
		Consider an M-winning coalition rule $\mathcal C$ and any sequential unanimity rule $\mathcal S$, whose associated sequence $\textrm S$ is an output of Algorithm 1 with M-winning coalition set $\mathcal C$ as input. For all profiles $R_N\in \mathcal R^N$, we have $\mathcal C(R_N) = \mathcal S(R_N)$. 
		\label{th:Alg1}
	\end{theorem}
	
	\noindent {\bf Proof.} Consider a profile $R_N$ and suppose that $C \in \mathcal C$ is such that $R_i=x$ for all $i \in C$, for some $x \in A$. Hence, $\mathcal C(R_N)=x$.
	
	Consider the sequence $\textrm S=(S_1,...,S_K)$ produced by Algorithm 1 with $\mathcal C$ as input. By construction, there exists $k \in \{1,...,K\}$ such that $S_k \subset C$. We distinguish between two cases: (i) $C=S_k$ and (ii) $S_k \subsetneq C$. In case (i), it follows that $S_l \in \mathcal C$ for all $l=1,...,k-1$. Hence, Moulin's property implies $S_l \cap C \neq \emptyset$ for all $l=1,...,k-1$. In case (ii), for any $l=1,...,k-1$, either $S_l \in \mathcal C$, or $S_l \not \in \mathcal C$. If $S_l \in \mathcal C$, then Moulin's property ensures $S_l \cap C \neq \emptyset$. If $S_l \not \in \mathcal C$, then condition (c) in Step 2 of Algorithm 1 implies $S_l \cap C \neq \emptyset$.
	
	Thus, for all $l<k$, there exists $i_l \in S_l \cap C$. Moreover, recall that $R_i=x$ for all $i \in C \supset S_k$. Hence, $\mathcal S(R_N)=x$. \qed

	\medskip
	
	\paragraph{An example involving non-weighted strong simple games.} We exhibit an application of Algorithm 1 to an M-winning coalition rule that corresponds to a non-weighted strong simple game. Recall the M-winning coalition set $\mathcal C^2$ in Section 2, i.e. 
	\begin{eqnarray*}
		\mathcal C^2 & = & \big \{ \{1,2,3\}, \{1,2,4\}, \{1,2,7\}, \{2,3,4\},\{1,3,5,6\},\{1,3,5,7\}, \{1,3,6,7\}, \{1,4, 5,6,8\} \\ 
		&&  \{2,3,5\},\{2,3,6\}, \{2,3,8\}, \{2,5,7\},\{2,6,7\},  \{2,4,5,6\}, \{3,4,7\} \big \},
	\end{eqnarray*}
	which we demonstrated to correspond to a non-weighted strong simple game. In applying Algorithm 1 to $\mathcal C^2$, we may begin by selecting voter 5 as the backstop, setting $N_1=\{5\}$. Subsequently, the only subset satisfying the requisite three properties is $\{1,2\}$ and so we set $N_2=\{1,2\}$. In the next iteration, the algorithm terminates. The remaining elements of $\mathcal C^2$ define the set $\mathcal C^*$:
	\begin{eqnarray*}
		\mathcal C^* & = & 		\big \{ \{2,3,4\},\{1,3,5,6\},\{1,3,5,7\}, \{1,3,6,7\}, \{2,3,6\}, \{2,3,8\}, \{2,6,7\}, \{3,4,7\} \big \}.
	\end{eqnarray*}
	Ordering the elements of $\mathcal C^*$ arbitrarily, we obtain the following sequence:
	\begin{eqnarray*}
		\textrm S & = & 		\big (\{1,3,6,7\}, \{2,3,6\}, \{2,3,4\}, \{3,4,7\}, \{1,3,5,6\},\{1,3,5,7\}, \{2,3,8\}, \{2,6,7\}, \{1,2\}, \{5\} \big ).
	\end{eqnarray*}

	Another application of Algorithm 1 to $\mathcal C^2$ begins with voter 8 as the backstop, setting $N_1=\{8\}$. Subsequently, it selects $N_2= \{1,3\}, N_3=\{1,2\}, N_4=\{2,5,6\}$. In the next (fifth) iteration, the algorithm terminates. The remaining elements of $\mathcal C^2$ define the set $\mathcal C^*$:
	\begin{eqnarray*}
		\mathcal C^* & = & 		\big \{ \{2,3,4\},\{2,3,5\},\{2,3,6\}, \{2,5,7\},\{2,6,7\}, \{3,4,7\} \big \}.
	\end{eqnarray*} 
	Ordering the elements of $\mathcal C^*$ arbitrarily, we obtain the following sequence:
	\begin{eqnarray*}
		\textrm S' & = & 		\big ( \{2,3,6\}, \{2,3,4\}, \{3,4,7\}, \{2,6,7\},\{2,3,5\},\{2,5,7\}, \{2,5,6\}, \{1,2\}, \{1,3\}, \{8\} \big ).
	\end{eqnarray*}
	
	\medskip
	
	The sequential unanimity rules associated with sequences S and $\textrm S'$ represent modest, but arguably nontrivial, ways of simplifying the strong simple game $(N,\mathcal C^2)$.
	
	\medskip
	
	A corollary of Proposition~\ref{prop:seq} and Theorem \ref{th:Alg1} is a new characterization of neutral and strategyproof rules.
	
	\begin{coro}
		A rule is strategy-proof and neutral if and only if it is a sequential unanimity rule produced via Algorithm 1.\label{coro:seq}
	\end{coro}
	\noindent{\bf Proof.} By Proposition \ref{prop:seq} all sequential unanimity rules are neutral and strategy-proof.
	Conversely, suppose rule $f$ is neutral and strategy-proof. By Theorem~\ref{th:moulin}, there exists some M-winning coalition set $\mathcal C^f$ and an associated M-winning coalition rule $\mathcal C^f$ such that $f=\mathcal C^f$. By Theorem \ref{th:Alg1} there exists a sequential unanimity rule $\mathcal S^f$ produced via Algorithm 1 such that $\mathcal C^f(R_N)=\mathcal S^f(R_N)$ for all profiles $R_N$. \qed

	\subsubsection{Essential sequential unanimity rules}
	
	In this section, we discuss a subfamily of sequential unanimity rules that are inspired by the proviso of Algorithm 1 and the efficiency justification behind it. 
	
	Due to the proviso, we know that the sequences produced by Algorithm 1 will never include two subsets $S_l, S_k$ such that $S_l \subsetneq S_k$ and $l<k$. By doing so, we exclude an obvious source of inefficiency in the implementation of the associated rule. Let us elaborate. If sequence S contained two such subsets $S_l$ and $S_k$ then, (i) if all voters in $S_k$ are unanimous, then so are all voters in $S_l$ and (ii) if there is disagreement among voters in $S_l$, then there is disagreements among those in $S_k$. Hence, subset $S_k$ would be superfluous to the implementation of the corresponding sequential unanimity rule. More precisely, for any profile $R_N$ the output of a sequential unanimity rule associated with S would be the same to the outcome of a similar rule in which subset $S_k$ has been removed from the sequence.\footnote{Note that this possibility is not purely theoretical. A valid application of Algorithm 1 to $\mathcal C^2$ of section 3.1.3 without the proviso produces the subset selections $N_1=\{8\}, N_2=\{1,3,6\}, N_3= \{1,3\}, N_4=\{1,2\}$ and sequence$$\textrm S=\big ( \{2,3,6\}, \{2,3,4\}, \{3,4,7\}, \{2,6,7\},\{2,3,5\},\{2,5,7\}, \{2,5,6\}, \{1,2\}, \{1,3\}, \{1,3,6\}, \{8\} \big ).$$Here, the penultimate element of the sequence S is superfluous to the associated sequential unanimity rule.}
	
	Hence, by imposing the proviso in Algorithm 1, we exclude an obvious kind of rule inefficiency due to two subsets in the sequence S containing each other. However, we do not know if the proviso guarantees efficiency in a broader sense. In particular, we do not know whether Algorithm 1 as specified will always produce sequences that contain no superfluous elements. Exploring this question is the motivation for the present subsection.

	Some definitions and notation are now necessary. {\em Essential} sequential unanimity rules are distinguished by the fact that every term in the associated sequence $\textrm S=(S_1,...,S_K)$ matters. That is, if we take out any subset $S_k$ from $\textrm S$ and consider the resulting sequential unanimity rule, that rule is {\em not} identical to the original sequential unanimity rule. In other words, there are no superfluous,``dummy" subsets in the sequence $\textrm S$. Note that $k$ can assume any value in $\{1,...,K-1\}$. The value $k=K$ is excluded because taking out the backstop voter results in a rule that is not well-defined. 
	
	\begin{Defi}
		A sequential unanimity rule $\mathcal S$ associated with sequence $\textrm S=(S_1,...,S_K)$ is {\bf essential} if, for all $k=1,...,K-1$, the sequential unanimity rule $\mathcal S_{-k}$ associated with sequence $\textrm S_{-k}= (S_1,...,S_{k-1}, S_{k+1},...,S_K)$ is such that there exists a profile $R_N \in \mathcal R^N$ such that $\mathcal S(R_N) \neq \mathcal S_{-k}(R_N)$.
	\end{Defi}
	
	Note that, given an $M$-winning coalition set $\mathcal C=\{C_1,..., C_L\}$, a sequence $\textrm S= (C_{j_1},...,C_{j_L}, \{i\})$, where $C_{j_1},...,C_{j_L}$ is an arbitrary ordering of $\{C_1,..., C_L\}$ and $i \in N$, cannot be associated with an essential sequential unanimity rule. This is because we can safely take out from $\textrm S$ any $C \in \mathcal C$ containing $i$ and the resulting rule will be identical to the original one. 
	
	We have not been able to ascertain whether the sequences produced via Algorithm 1 lead to essential sequential unanimity rules. However, what we can say with certainty is that if a fourth condition is added to the criteria of step 2 of Algorithm 1, then the procedure produces essential sequential unanimity rules.
	
Before stating this condition, we need some further notation. At each iteration $k$ of Algorithm 1, define $\mathcal E_k= \{C \in \mathcal C_{k-1}: \; N_k \subsetneq C\}$. Thus, $\mathcal E_k$ is the set of winning coalitions within $\mathcal C_{k-1}$ that are proper supersets of the newly selected subset $N_k$. As such, it coincides with the set of winning coalitions that is taken out of $\mathcal C_{k-1}$ at time $k$, in order to update it to $\mathcal C_k$. 
	
	The additional condition to step 2 of Algorithm 1, which ensures that the produced sequential unanimity rules are essential, is as follows.
	\begin{itemize}
		\item [2(iv)] If $k \geq 3$, then $N_k \not \subset C$ for all $C \in \bigcup \limits_{l=2}^{k-1} \mathcal E_{l} \equiv \left (\mathcal C_{k-1}\setminus \mathcal C_1 \right)$ .
	\end{itemize}
	
	Notice that, as Algorithm 1 proceeds and $k$ increases, set $\mathcal C_{k-1}$ grows and condition 2 (iv) becomes harder to satisfy. 

	We are now ready to prove the main result of this sub-section.

	\begin{prop} Consider an M-winning coalition rule $\mathcal C$ and any sequential unanimity rule $\mathcal S$, whose associated sequence $\textrm S$ is an output of Algorithm 1 satisfying additional condition 2 (iv), with M-winning coalition set $\mathcal C$ as input. The sequential unanimity rule $\mathcal S$ is essential. \label{prop:esse}
	\end{prop}
	\noindent {\bf Proof.} Let $\textrm S=(S_1,...,S_K)$ be an output of Algorithm 1 with $\mathcal C$ as input. We assume that $K>1$ and that, at each iteration of Algorithm 1, condition 2 (iv) is satisfied. Let $\mathcal S$ denote the corresponding sequential unanimity rule. Suppose subset $S_k$ is removed from sequence $\textrm S$ for some $k=1,...,K-1$. The updated sequential unanimity rule $\mathcal S_{-k}$ is associated with sequence $(S_1,...,S_{k-1},S_{k+1},...,S_K)$. 
	
%
	Let $k=1,...,K-1$ and $C \in \mathcal E_k$. We consider two cases:
	
\noindent{\bf Case 1:} $k=1$. Let $R_N$ be a profile such that $R_i=a$ if and only if $i \in C$. Then, $\mathcal S(R_N)=a$. Let $l =2,...,K$. If $S_l \subsetneq C$, then $C \in \mathcal C_l$, which is a contradiction. Thus, there exists $i_l \in S_l \setminus C$, implying $R_{i_l}=b$. Since $S_K$ is a singleton, by Definition \ref{def:sur}, $S_{-1}(R_N)=b$.
		
		If $K=2$ we are done. So suppose that $K>2$.

		\noindent{\bf Case 2:} $k=2,...,K-1$. Let $R_N$ be a profile such that $R_i=a$ if and only if $i \in C$. Let $l=1,...,k-1$. By condition 2(iv), there exists $i_l \in S_l \setminus C$, implying $R_{i_l}=b$. By condition 2(iii) of Algorithm 1, there exists $j_l \in S_l \cap C$. Thus, $R_{j_l}=a$. By Definition \ref{def:sur}, $\mathcal S(R_N)=a$. Let $l=k+1,...,K$. By a similar argument as in Case 1, there exists $i_l \in S_l \setminus C$, implying $R_{i_l}=b$. 
		
		Thus, for all $l=1,...,K$ such that $l \neq k$, there exists $i_l \in S_l$ such that $R_{i_l}=b$. Since $S_K$ is a singleton, by Definition \ref{def:sur}, $S_{-1}(R_N)=b$.

	\qed

	\medskip

	Before concluding this section, it is worth noting that while condition (iv) is sufficient to guarantee that sequential unanimity rules will be essential, we do not know if imposing it in Algorithm 1 is necessary to achieving this goal. In fact, we do not know if {\em any} other condition is needed beyond the proviso, since condition (iv) is automatically satisfied in all of the applications of Algorithm 1 that appear in this paper. 
	

%
%
%

	\subsection{From sequential unanimity rules to M-winning coalition rules}
	
	In this section, we move in the reverse direction and demonstrate that a sequential unanimity rule can be transformed into an equivalent M-winning coalition rule. Our approach is once again constructive, and we refer to the algorithm that we develop and analyze as {\bf Algorithm 2}.
	
	\subsubsection{Algorithm 2: A simple special case}
	
	To aid the reader, we begin by describing how Algorithm 2 works for sequential unanimity rules with associated sequences $\mathrm S=(S_1,...,S_K)$ such that $S_l \cap S_k = \emptyset$, for all $k,l \in \{1,...,K\}$ where $k\neq l$. Thus, we focus on the special case in which each voter can appear at most once within the sequence S.

	
	Algorithm 2 begins by defining a candidate winning coalition set $\mathcal C_1$ and initializing it to $\mathcal C_1= S_1$. This implies that the sequential unanimity rule $\mathcal S$ and any M-winning coalition rule containing the winning coalition of $\mathcal C_1$ produce the same outcome for all profiles featuring unanimous agreement within subset $S_1$. Such profiles ensure that the sequential unanimity rule $\mathcal S$ reaches an outcome immediately after consulting the first element of the sequence S, i.e., $S_1$.
	
	Susequently, the algorithm considers subset $S_2$ and does the following. For every $i \in S_1$, it adds the subset $S_2 \cup \{i\}$ to winning coalition set $\mathcal C_1$, updating the latter to $\mathcal C_2$. This ensures that rule $\mathcal S$ and any M-winning coalition rule containing the winning coalitions of $\mathcal C_2$ produce the same outcome for all profiles featuring (i) unanimous agreement within subset $S_1$, or (ii) some voter disagreement within subset $S_1$, followed by  unanimous agreement within subset $S_2$. These profiles are such that the rule $\mathcal S$ reaches an outcome after consulting at most the first two elements of sequence S.

	Algorithm 2 continues iteratively in this fashion for all $k=3,...,K$. That is, at each iteration $k$ it does the following: For every $(i_1,i_2,...,i_{k-1}) \in S_1 \times S_2 \times .... S_{k-1}$, the subset $\{i_1,i_2,...,i_{k-1}\} \cup S_k$ is added to winning coalition set $\mathcal C_{k-1}$, updating the latter to $\mathcal C_k$. As a result, rule $\mathcal S$ and any M-winning coalition rule containing the winning coalitions of $\mathcal C_k$ produce the same outcome for all profiles featuring  (i) unanimous agreement within subset $S_1$, or (ii) for some $l\in \{2,...,k\}$,  some voter disagreement within each subset $S_1, S_2,...,S_{l-1}$, followed by unanimous agreement within subset $S_l$. These profiles are such that the rule $\mathcal S$ reaches an outcome after consulting at most the first $k$ elements of sequence S.
	
	Algorithm 2 terminates after iteration $K$, setting $\mathcal C^S=\mathcal C_K$. At that point, it is possible to show that the set $\mathcal C^S$ is an M-winning coalition set as per Definition \ref{def:wcs}. Moreover, the associated M-winning coalition rule $\mathcal C^S$ satisfies $\mathcal C^S(R_N)= \mathcal S(R_N)$ for all $R_N \in \mathcal R_N$.

	\subsubsection{Paths between subsets}

	When the sequence S features overlapping subsets, the previous description of Algorithm 2 is very inefficient. This is because many subsets of the form $\{i_1,i_2,...,i_{k-1}\} \cup S_k$ will be proper supersets of other similar subsets. As a result, they will need to be removed from the final winning set $\mathcal C^S$ to ensure that the latter is an M-winning coalition set. Considering that the number of subsets of the form $\{i_1,i_2,...,i_{k-1}\} \cup S_k$ grows exponentially in $k$, it would be very inefficient for our algorithm to indiscriminately add all such subsets into a candidate set of winning coalitions, only to remove a majority of them afterwards.
	
	This concern motivates the introduction of a new concept. Given a sequence $\textrm S=(S_1,...,S_K)$ of subsets of $N$, we introduce the notion of a {\bf path} between $S_l$ and $S_k$, for $l,k \in \{1,..,K\}$ and $l<k$. Intuitively, a sequence of voters $(i_1, i_2,....)$ constitutes a {\bf path} between $S_l$ and $S_k$ if it represents a parsimonious traversal of the sequence $(S_l, S_{l+1},,...,S_k)$. What do we mean by parsimonious? Essentially we mean that, when selecting voters sequentially from $S_l$ to $S_k$, if a voter is chosen from some subset $S_h$, then all successive subsets of $\textrm S$ in which that voter appears need not be considered en route to $S_k$. Thus, if voter $i_l \in S_l$ is selected, then it is as if all successive subsets containing voter $i_l$ have already been visited, and can be disregarded as we make our way to  from $S_l$ to $S_k$. A formal definition follows.

	\begin{Defi}Consider a sequence $(S_1,..., S_K)$ of non-empty subsets of $N$.\footnote{Note that, unlike in the sequences used for sequential unanimity rules, we make no restrictions on $\textrm S$.} Given $l,k \in \{1,..,K\}$ such that $l<k$, a {\bf path $p$} from $S_l$ to $S_k$ is a finite sequence $p=(i_1, i_2...)$ of voters constructed via the following algorithm:
		
		\begin{itemize}
			\item [0.] Let $\textrm S=(S_l,...,S_k)$. Let $S^0=(S^0_1, S^0_2,...)$ denote the sequence obtained from S by removing all $S_j$ such that $S_j \cap S_k \neq \emptyset$. 
			\item [1.] If sequence $S^{0}$ is empty (i.e., has length zero), STOP. There are no paths from $S_l$ to $S_k$. Otherwise, select $i_1 \in S^0_1$. Delete all sets $S_j$ from the sequence $S^0$ such that $i_1 \in S_j$. Denote the updated sequence $S^1=(S^1_1, S^1_2,...)$.
			\item [2.] For $h = 2,3,...$
			\begin{itemize}
				\item [a.] If $S^{h-1}$ is empty, STOP. The computed path is $p=(i_1, i_2,...,i_{h-1})$.
				\item [b.] Otherwise, pick $i_h \in S^{h-1}_1$. Delete all sets $S^{h-1}_j$ from the sequence $S^{h-1}$ such that $i_h \in S^{h-1}_j$. Denote the updated sequence $S^h=(S^h_1,S^h_2,...)$.
			\end{itemize}
		\end{itemize}
		\label{def:path}
	\end{Defi}
	
	Given a sequence $(S_1,...,S_K)$ there will generally exist a number of paths between any two subsets $S_l$ to $S_k$. Consider the following example.
	
	\paragraph{Example 3.} Suppose $S_1=\{1,2,3\}, S_2=\{6,7\}, S_3=\{1,2,4\}, S_4=\{2,5,6\}, S_5=\{2,8\}, S_6=\{8\}$. Let $(S_1, S_2,S_3,S_4, S_5, S_6)$ and suppose we want to identify all paths from $S_1$ to $S_6$. Applying Definition~\ref{def:path}, we calculate 16 such paths from $S_1$ to $S_6$. Here they are: $(1,6), (1,7,2), (1,7,5), (1,7,6), (2,6), (2,7), (3,6,1), \\ (3,6,2), (3,6,4), (3,7,1,2), (3,7,1,5), (3,7,1,6), (3,7,2), (3,7,4,2), (3,7,4,5), (3,7,4,6)$. 
	
\begin{figure}[H]
		\centering
		\includegraphics[height=.35\textheight, width=.9\textwidth]{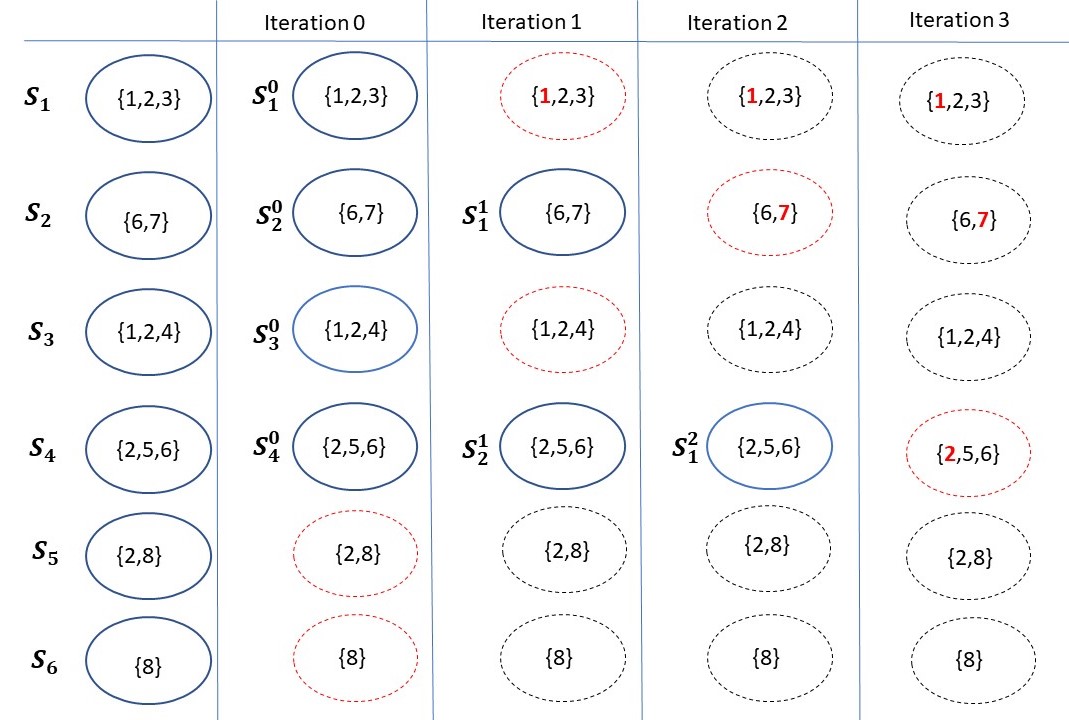}
		\caption{Example 3: Computation of path (1,7,2). Deleted subsets appear in dashed circles. Newly deleted subsets appear in red dashed circles. Selected voters are indicated in red, bold fonts.}
		\label{fig:path_Ex1}
	\end{figure}

	Figure~\ref{fig:path_Ex1} illustrates the computation of path $(1,7,2)$. We initialize the process by deleting all elements of sequence $\textrm S$ that have a non-empty intersection with $S_6$, i.e., that contain the voter 8. Thus, $S_5$ and $S_6$ are deleted and the resulting sequence is denoted $S^0=(S_1^0, S_2^0, S_3^0, S_4^0)=(\{1,2,3\},\{6,7\},\{1,2,4\}, \{2,5,6\})$. In iteration 1, we select $i_1=1 \in S_1^0$. Subsequently, we delete all the terms of sequence $\textrm S^0$ that contain voter 1. This means we delete sets $S_1^0, S_3^0$. The resulting sequence is denoted $S^1=(S_1^1, S_2^1)=(\{6,7\},\{2,5,6\})$. In iteration 2, we select $i_2=7 \in S_1^1$. Subsequently, we delete all the terms of sequence $\textrm S^1$ that contain voter 7. This means we delete set $S_1^1$. The resulting sequence is denoted $\textrm S^2=(S_1^2)=(\{2,5,6\})$. In iteration 3, we select voter $i_3=2 \in S_1^2$. Subsequently, we delete set $S_1^2$ and obtain an empty sequence, which we denote by $S^3$. In iteration 4, the algorithm terminates with the computed path $(i_1, i_2, i_3)=(1,7,2)$.
	
	%
	\subsubsection{General description of Algorithm 2}
	
	At a high level, Algorithm 2 works in the following way. Given a sequence of subsets $\textrm S=(S_1,...,S_K)$, it begins by putting $S_1$ into a candidate winning coalition set $\mathcal C_1$. At each subsequent iteration, the algorithm adds further elements to this set, each time updating it. Namely, at iteration $k=2,...,K$, Algorithm 2 considers subset $S_k$ of sequence $\textrm S$, and does the following:
	\begin{itemize}
		\item[(i)] It computes all paths from $S_1$ to $S_k$; and
		\item [(ii)] For each path $p$ from $S_1$ to $S_k$, the subset $\{p\} \cup S_k$ is added to the candidate winning coalition set. We denote the {\em collection} of all such subsets by $\mathcal C^p_k$, i.e. $\mathcal C^p_k=\bigcup \limits_{p}\left\{ \{p\} \cup S_k\right\}$. Correspondingly, set $\mathcal C_{k-1}$ is updated to $\mathcal C_k = \mathcal C_{k-1} \cup \mathcal C^p_k$. If necessary, set $\mathcal C_k$ is pruned in order to get rid of elements which are proper supersets of others.
	\end{itemize}
	
	After the last iteration $k=K$, the candidate winning coalition set is $\mathcal C_{K}$. At that point, the algorithm terminates and produces as output the M-winning coalition set $\mathcal C^{\textrm S}=\mathcal C_K$. 
	
	
	At each iteration $k$, rule $\mathcal S$ and any M-winning coalition rule containing the winning coalitions in $\mathcal C_k$ produce the same outcome for all profiles featuring (i) unanimous agreement within subset $S_1$, or, (ii) for some $l \in \{2,...,k-1\}$ some voter disagreement within each subset $S_1, S_2, .... S_{l-1}$, followed by unanimous agreement within subset $S_l$. These profiles are such that the sequential unanimity rule $\mathcal S$ reaches an outcome after consulting at most the first $k$ elements of sequence S.
	

	\begin{figure}[H]
		\centering
		\includegraphics[height=.15\textheight, width=.67\textwidth]{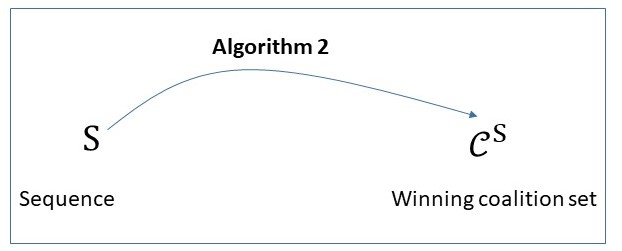}
		\caption{Algorithm 2 associates to each sequence $\textrm S$ a winning coalition set $\mathcal C^{\textrm S}$.} 
	\label{fig:Alg2}
\end{figure}

We will show two things: (i) if $\textrm S=(S_1,...,S_K)$ is such that $|S_K|=1$, then the winning coalition set $\mathcal C^{\textrm S}$ produced by Algorithm 2 is an M-winning coalition set as per Definition~\ref{def:wcs}, and (ii) the sequential unanimity rule $\mathcal S$ with associated sequence $\textrm S$, is equivalent to the M-winning coalition rule $\mathcal C^{\mathrm S}$. Before proceeding with these results, we illustrate Algorithm 2 on the usual job-market candidate example.

\subsubsection{An illustrative example}

Recall the winning coalition set $\mathcal C$ depicted in Figure \ref{fig:Alg1_1}. In the previous chapter, we showed that by running Algorithm 1 with $\mathcal C$ as input in two different ways we obtained the sequences $$\textrm S^1= (\{1,2\}, \{3,4\}, \{5,6\}, \{7\})$$and$$\textrm S^2=(\{1,2\}, \{3,5,7\}, \{3,6,7\}, \{3,5,6\},\{4\}).$$By Theorem~\ref{th:Alg1} the respective sequential unanimity rules are equivalent to the M-winning coalition rule $\mathcal C$. For the results of this section to be consistent with Theorem~\ref{th:Alg1}, when applying Algorithm 2 to sequences $\textrm S^1$ and $\textrm S^2$ we need to obtain the same M-winning coalition set $\mathcal C$ of Figure \ref{fig:Alg1_1}. We thus verify that this is the case.

First, consider $\textrm S^1$. Since this sequence features completely non-overlapping subsets, path calculations are very easy. Indeed, given $k\in \{2,3,4\}$, the set of paths from $S_1$ to $S_k$ will consist of all sequences $(i_1,...,i_{k-1})$ that satisfy $i_1 \in S_1, ..., i_{k-1} \in S_{k-1}$. The corresponding additions to the candidate winning coalition set are straightforward. Finally, after iteration $K$, we verify that $\mathcal C_K = \mathcal C^S$ equals the M-winning coalition set of Figure~\ref{fig:Alg1_1}.

\begin{figure}[H]
	\centering
	\includegraphics[height=.35\textheight, width=.9\textwidth]{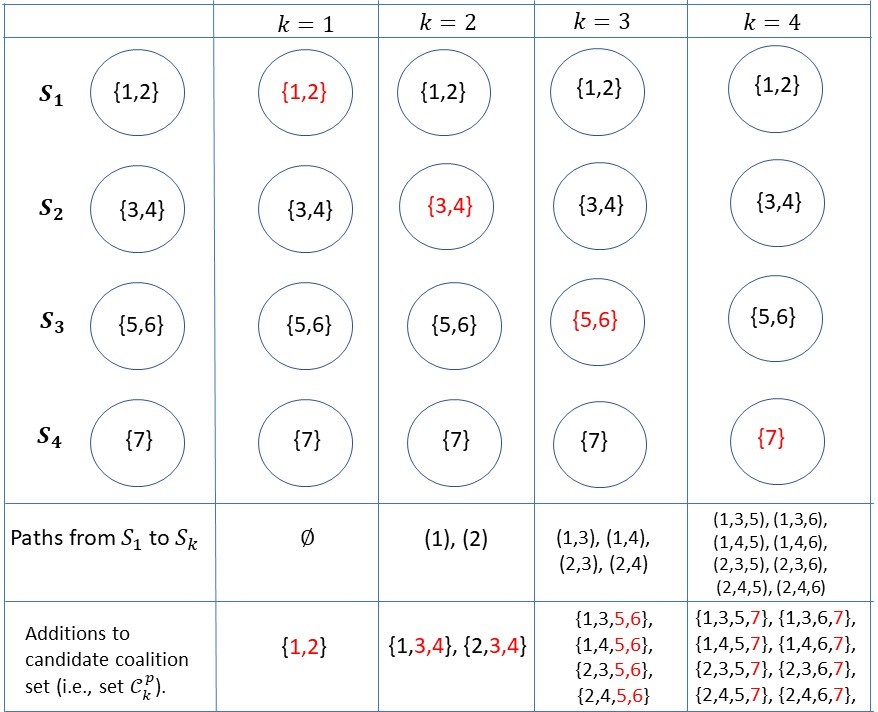}
	\caption{Applying Algorithm 2 to $\textrm S^1=(\{1,2\},\{3,4\},\{5,6\},\{7\})$. We verify that $\mathcal C_4=\mathcal C^S$ equals the M-winning coalition set of Figure~\ref{fig:Alg1_1}. }
	\label{fig:Alg2_Ex1}
\end{figure}

\begin{figure}[H]
	\centering
	\includegraphics[height=.4\textheight, width=.99\textwidth]{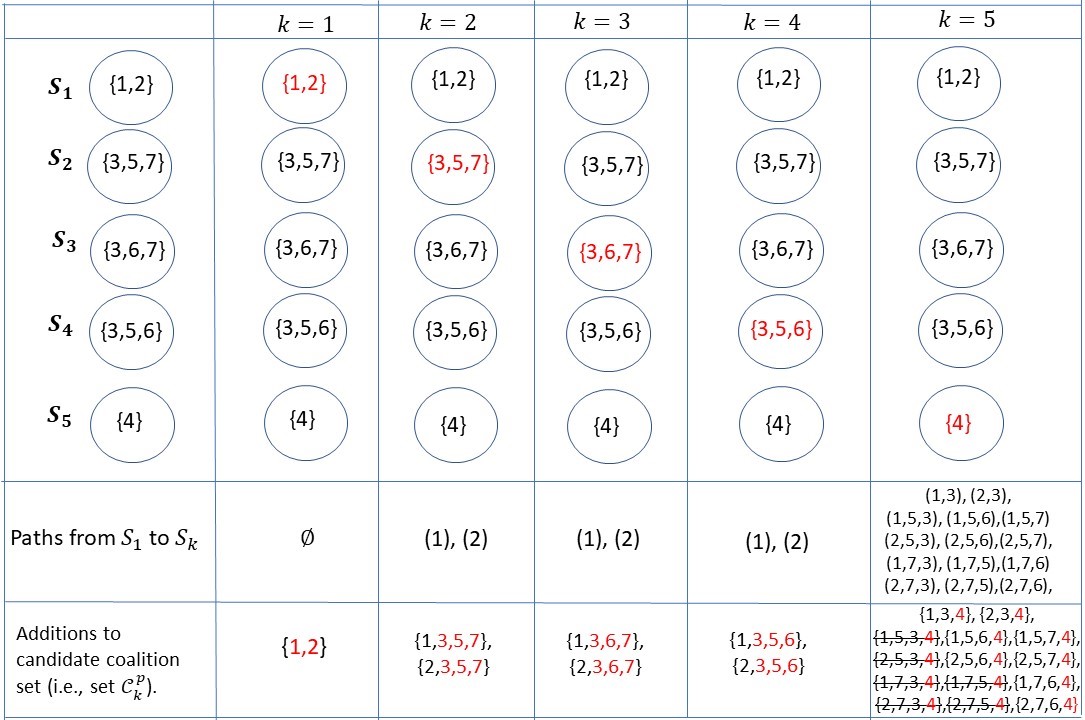}
	\caption{Applying Algorithm 2 to $\textrm S^2=(\{1,2\},\{3,5,7\},\{3,6,7\}, \{3,5,6\}, \{4\})$. Subsets which are removed during the pruning phase of the Algorithm, because they are strict supersets of others, are crossed out (e.g. $\{1,5,3,4\} \supsetneq \{1,3,4\}$). Duplicate subsets are crossed out as well (e.g. $\{2,7,5,4\}$ is identical to $\{2,5,7,4\}$). The resulting set $\mathcal C_5= \mathcal C^S$ equals the M-winning coalition set of Figure~\ref{fig:Alg1_1}.}
	\label{fig:Alg2_2}
\end{figure}

The application of Algorithm 2 is a little more involved for $\textrm S^2$. Here, there is overlap between subsets $S_2, S_3, S_4$ so that path calculations are a little trickier. Figure~\ref{fig:Alg2_2} specifies at each iteration of the algorithm, the paths and additions to the candidate coalition set. At iteration 5, four subsets need to be removed from $\mathcal C_K$ because they are proper supersets of others.

\subsubsection{Algorithm 2 and its properties}

In this section we formally define Algorithm 2 and prove its desirable properties.

\begin{center}
	\fbox{\begin{minipage}{\linewidth}
			\noindent {\bf Algorithm 2}$\big |$ \; \;  {\bf Input}: Sequence $\textrm S=(S_1,...,S_K)$. {\bf Output}: winning coalition set  $\mathcal C^{\textrm S}$
			\hrule
			\begin{itemize}
				\item [1.] Let $\mathcal C_1= S_1$.
				\item [2.] For $k=2,...,K$
				\begin{itemize}
					\item [a.]  Consider the set $S_k$ and the sequence $(S_1, ...., S_k)$.
					\item [b.] For every path $p$ from $S_1$ to $ S_k$, define $C_p \subset N$ such that $C_p = \{p\} \cup  S_k$. Let $\mathcal C^p_k=\bigcup \limits_{p}\left\{ \{p\} \cup S_k\right\}$ be the collection of all such subsets $C_p$. If no paths from $S_1$ to $S_k$ exist, set $\mathcal C^p_k = S_k$.
					\item [c.] Define $\mathcal{C}_k= \mathcal C_{k-1} \cup \mathcal C^p_k$. Update $\mathcal C_k$ by deleting all $\tilde C \in \mathcal{C}_{k}$ satisfying $\tilde C \supsetneq C$ for some $C \in \mathcal{C}_{k}.$ 
				\end{itemize}
				\item [3.] Define $\mathcal C^S\equiv \mathcal C_K$.
			\end{itemize}
	\end{minipage}}
\end{center}

\bigskip

\begin{lemma}
	Let $\textrm S=(S_1,...,S_K)$ be a sequence of subsets of $N$ such that $|S_K|=1$. The winning coalition set $\mathcal C^{S}$ computed by Algorithm 2 is an M-winning coalition set.\label{lem:Mwin}
\end{lemma}

\noindent {\bf Proof.} By Step 2 (iii) of the algorithm, the set $\mathcal C^S$ satisfies property P1 (Minimality). We show that $\mathcal{C}^S$ satisfies property P2 (Moulin's property). First, we show that for all $C, C'\in \mathcal{C}^S$, $C\cap C'\neq \emptyset$. Let $C, C'\in \mathcal{C}^S$. By construction, there exist $k,l \in \{1,...,K\}$ and two paths $ p$ and $p'$ from $S_1$ to $S_k$, from $S_1$ to $S_\ell$, respectively, such that $C=S_k\cup \{p\}$ and $C'=S_\ell \cup \{p'\}$. If $k=\ell$, then $S_k=S_\ell$, and thus $C\cap C'\neq \emptyset$. Suppose that $\ell<k$. If $S_l \cap S_k \neq \emptyset$, then $C\cap C'\neq\emptyset$.  If $S_l \cap S_k = \emptyset$, by construction there is an element $i$ of the path $p$ such that $i\in S_\ell$. Hence, $\{p\} \cap S_\ell \neq \emptyset$. Therefore, $C\cap C'\neq\emptyset$.

Finally, let $N'$ be a subset of $N$ such that for all $C\in \mathcal{C}^S$, $N'\cap C\neq \emptyset$. We show that there exists $C \in \mathcal{C}^S$ such that $C'\subset N'$. Let us first prove a claim.	
\paragraph{Claim 1.} {\em There exists $k=1,...,K$ such that $S_k\subset N'$.}

\medskip

\noindent{\bf Proof:} We proceed by contradiction. Suppose that for all $k=1,...,K$, there is $i_k\in S_k$ such that $i_k\notin N'$. By construction, there exists a path $p$ from $S_1$ to $S_K$ that is a subsequence of $(i_1,...,i_{K-1})$. Moreover, since $S_K$ is a singleton, then $S_K\cap N'=\emptyset$. By construction, the set $S_K\cup \{p\}$, or a proper subset of it, will belong to $\mathcal{C}^S$. Let us call this subset $C'$. Since $N'\cap S_K=\emptyset$ and $N'\cap p=\emptyset$, we have identified a subset $C'\in \mathcal C^S$ such that $N'\cap C'=\emptyset$. This conclusion contradicts our assumption that for all $C\in \mathcal{C}^S$, $N'\cap C\neq\emptyset$. This concludes the proof of Claim 1.

\medskip

Thus, let us consider the index $k \in \{1,...,K\}$ such that $S_k\subset N'$. If $k=1$, then $S_1\subset N'$ and we are done since $S_1\in \mathcal{C}^S$. So, suppose that $k>1$. We prove the following claim. 

\paragraph{Claim 2.} {\em Let $k$ be the minimum index such that $S_k\subset C$. For all $\ell<k$, $S_\ell \cap N'\neq \emptyset$.}
\medskip

\noindent{\bf Proof:} We proceed by contradiction. Suppose that there is $\ell<k$ such that $S_\ell\cap N'=\emptyset$. If $\ell=1$, then we have $S_1\in \mathcal{C}^S$ and $N'\cap S_1=\emptyset$. This conclusion contradicts our assumption that for all $C\in \mathcal{C}^S$, $N'\cap C\neq \emptyset$. Suppose that $\ell>1$. Since $k$ is chosen to be minimum, for all $\ell'=1,...,\ell-1$, there is $i_{\ell'}\in S_{\ell'}$ such that $i_{\ell'}\notin N'$. Thus, there exists a path $p$ from $S_1$ to $S_\ell$ that is a subsequence of $(i_1,...,i_{\ell-1})$. By construction, the set $S_l\cup \{p\}$, or a proper subset of it, will belong to $\mathcal{C}^S$. Let us call this subset $C'$. Since $S_\ell\cap N'=\emptyset$ and $\{p\} \cap N'=\emptyset$, we have identified a subset $C'\in \mathcal C^S$ such that $N'\cap C'=\emptyset$. This conclusion contradicts our assumption that for all $C \in \mathcal{C}^S$, $N'\cap C \neq \emptyset$. This concludes the proof of Claim 2.

\medskip

In light of Claim 2, for all $\ell=1,...,k-1$, let $i_\ell\in S_\ell \cap N'$. Thus, there exists a path $p$ from $S_1$ to $S_k$ that is a subsequence of $(i_1,...,i_{k-1})$. By construction, the set $S_k\cup \{p\}$, or a proper subset of it, belongs to $\mathcal{C}^S$. We call this subset $C'$. Thus, we have identified $C' \in \mathcal C^S$ such that $C' \subset N'$. \qed

\medskip

\begin{theorem}
	Consider a sequential unanimity rule $\mathcal S$ parametrized by $S=(S_1,...,S_K)$. Suppose Algorithm 2 is applied to sequence $S$, producing the M-winning coalition set $\mathcal C^S$, and consider the associated M-winning coalition rule $\mathcal C^{S}$. For all profiles $R_N \in \mathcal R^N$, we have $\mathcal S(R_N) = \mathcal C^{S}(R_N)$.
	\label{th:Alg2}
\end{theorem}

\noindent {\bf Proof.} Consider a profile $R_N$. Suppose that at this profile set $S_k$ is decisive. That is, (i) there exists $x \in A$ such that $R_i=x$ for all $i \in S_k$ and (ii) for all $l < k$, there exist  $i,j, \in S_l$ such that $R_i \neq R_j$. Then $\mathcal S(R_N)=x$. We will show that $\mathcal C^S(R_N)=x$.

By Lemma~\ref{lem:Mwin}, $\mathcal C^S$ is an M-winning coalition set. Thus, we need to prove that there exists $C \in \mathcal C^S$ such that $R_i=x$ for all $i \in C$. Recall $k$, the index of the decisive set. If $k=1$, we immediately conclude $\mathcal C^S(R_N)=x$, since $S_1 \in \mathcal C^S$. If $k>1$, then for all $l <k$, there exists $i_l \in S_l$ such that $R_i=x$. Therefore, there exists a path $p^*$ from $S_1$ to $S_k$ that is a subsequence of $(i_1,...,i_{l-1})$. The corresponding set $S_k \cup \{p^*\}$ will either belong to $\mathcal C^S$, or step 2 iii of the algorithm will ensure that a strict subset of it does. Either way, $\mathcal C^S$ will include a set $C'$ such that $R_i=x$ for all $i \in C'$. Thus, $\mathcal C^S(R_N)=x$. \qed


\begin{coro}
	Suppose sequence $\textrm S$ is obtained by applying Algorithm 1 on M-winning coalition set $\mathcal C$. If we apply Algorithm 2 to $\textrm S$, we obtain $\mathcal C^{\textrm S} = \mathcal C$.
	\label{coro:12}
\end{coro}
\noindent {\bf Proof.} By Theorem~\ref{th:Alg1}, the M-winning coalition rule $\mathcal C$ is equivalent to the sequential unanimity rule $\mathcal S$ associated with $\textrm S$. This implies that when applying Algorithm 2 to $\textrm S$, we obtain a M-winning coalition set $\mathcal C^{\textrm S}$ whose implied M-winning coalition rule is equivalent to $\mathcal C$. Since a given neutral and SP rule is uniquely defined by its M-winning coalition set, we conclude $\mathcal C^{\textrm S} = \mathcal C$.  \qed

\section{Indifferences}

We now show how our results extend to the full preference domain, which allows voters to be indifferent between alternatives. The candidate set $A$ is expanded to include $0$, indicating a tie between $a$ and $b$. Voter $i$'s preference relation over candidates is denoted by $\tilde R_i \in \{a,b,0\}$. A preference of $\tilde R_i=x$, with $x \in \{a,b\}$, means that voter $i$ has a strict preference for candidate $x$, whereas $\tilde R_i=0$ means that $i$ is indifferent between $a$ and $b$. A voter with a strict preference for $x \in \{a,b\}$ is assumed to prefer the null outcome 0 to an outcome $y \in \{a,b\}$ such that $y \neq x$.

The property of strategy-proofness ensures that it is not possible for a voter to misreport his preferences and obtain an outcome that he prefers to that obtained under truthfulness. Since preferences in the full domain are a little more involved, the following definition clarifies what we mean.

\begin{Defi} \label{def:stpind}
A rule $f$ is {\bf strategy-proof} if for all profiles $\tilde R_N \in \mathcal{\tilde R}^N$, each voter $i \in N$, and each $\tilde R'_i \in \{a,b,0\}$ such that $\tilde R_i' \neq \tilde R_i$, $i$ does not prefer $f(\tilde R'_i,\tilde R_{-i})$ to $f(\tilde R_N)$.
\end{Defi}
Thus, to establish a rule's strategy-proofness it is sufficient to focus on voters having a strict preference for candidate $a$ or $b$. If such voters cannot profitably misreport their preferences, then the criterion is met.

	We need to take similar care in adapting the definition of neutrality. Given a permutation $\pi: \{a,b, 0\} \mapsto \{a,b,0\}$ such that $\pi (0)=0$ and a profile $\tilde R_N \in \mathcal{\tilde R}^N$, define the profile $\pi \tilde R_N$ as $\pi \tilde R_i = \pi (\tilde R_i)$ for all $i \in N$. Thus, voters who are indifferent between candidates remain indifferent after the identities of $a$ and $b$ have been flipped by $\pi$.

\begin{Defi} \label{def:neutind}
	A rule $f$  is {\bf neutral} if for each permutation $\pi$ of $ \{a,b,0\}$ such that $\pi(0)=0$, and all profiles $\tilde R_N \in \mathcal {\tilde R}^N$,
	\begin{equation*}
		f(\pi \tilde R_N)=\pi(f(\tilde R_N)).
	\end{equation*}
\end{Defi}


With indifferences, the M-winning coalition rules of Section 3 may not be well-defined. They thus need to be modified. Consider a neutral and strategy-proof rule $f: \mathcal{\tilde R}^N \mapsto \tilde \{a,b,0\}$. Examples of such a rule $f$ may include majority rule and dictatorship where, to be clear, majority rule is defined as the rule $f^M$ satisfying:
\begin{equation*}
	f^M(\tilde R_N)=
	\left\{\begin{array}{ccl}
		a, & &\text{if   } |i \in N: \; s.t. \; \tilde R_i=a| > |i \in N: \; s.t. \; \tilde R_i=b|\\
		b, & &\text{if   } |i \in N: \; s.t. \; \tilde R_i=b| > |i \in N: \; s.t. \; \tilde R_i=a|\\	
		0, & & \text{otherwise.}\\
	\end{array}\right.
	\label{eq:maj}
\end{equation*}

\begin{Defi}
	Consider an M-winning coalition set $\mathcal C$ and a neutral and strategy-proof rule $f: \mathcal{\tilde R}^N \mapsto \{a,b,0\}$. The {\bf M-winning coalition rule with $f$ as default}  $\mathcal C^{f}: \mathcal{\tilde R}^N \mapsto \{a,b,0\} $ is defined as follows: for all $\tilde R_N \in \mathcal{\tilde R}^N$ and $x \in \{a,b\}$,
	$$\exists C \in \mathcal C \textrm{  s.t.  } \tilde R_i=x \textrm{   for all  } i \in C \; \Rightarrow \; \mathcal C(\tilde R_N)=x;$$otherwise, $\mathcal C^f(\tilde R_N)=f(\tilde R_N).$
\end{Defi}

So, if there exists an element of $\mathcal C$ that unanimously supports candidate $x \in \{a,b\}$, the rule $\mathcal C^{f}$ picks it. If there is no such consensus, we move on to a second stage in which the outcome is determined via the default rule $f$ that retains the desirable features of strategy-proofness and neutrality. This two-stage way of redefining M-winning coalition rules to accommodate the full domain is consistent with Bartholdi et al.'s~\cite{b21} treatment. Now, we move on to sequential unanimity rules. The full-domain adaptation here is a little more subtle.


\begin{Defi}
	Consider a sequence $S=(S_1,...,S_K)$ of subsets of $N$ satisfying $|S_K|$=1 and a neutral and strategy-proof rule $f: \mathcal{\tilde R}^N \mapsto \{a,b,0\}$. The {\bf sequential unanimity rule with $f$ as default} $\mathcal S^f: \mathcal{\tilde R}^N \mapsto \{a,b,0\}$, is defined as follows:
	\begin{itemize}
		\item [1.] If $\tilde R_i=x \in \{a,b\}$ all $i \in S_1$, then $\mathcal S^f(\tilde R_N)=x$. Otherwise
		\item [2.] For $k=2,...,K$
		\begin{itemize}
			\item If $\tilde R_i=x\in \{a,b\}$ for all $i \in S_k$, and for all $l=1,...,k-1$ there exists $i_l \in S_l$ such that $R_{i_l}=x$, then STOP. We have $\mathcal S^f(\tilde R_N)=x$.
			\item If $\tilde R_i=0$ for all $i \in S_k$, then STOP. We have $\mathcal S^f(\tilde R_N)=f(\tilde R_N)$.
		\end{itemize}
		\item [3.] If the above procedure reaches $k=K$ without stopping, then $\mathcal S^f(\tilde R_N)=f(\tilde R_N)$.
	\end{itemize}
\end{Defi}

\medskip

The adapted versions of M-winning coalition and sequential unanimity rules retain their desirable features. The next Proposition states this result.

\begin{prop}
	Let $f: \mathcal{\tilde R}^N \mapsto \{a,b,0\}$ be a neutral and strategy-proof rule. Then, all M-winning coalition rules and all sequential unanimity rules with $f$ as default are neutral and strategy-proof.
\end{prop}
\noindent{\bf Proof.} It is easy to show that M-winning coalition rules with $f$ as default are neutral. We turn to strategy-proofness. Consider a rule $\mathcal C^f$ with associated M-winning coalition set $\mathcal C$ and neutral and default $f$. Let $\tilde R_N$ be a profile. We consider two cases: (i) there exist $C \in \mathcal C$ and $x \in \{a,b\}$ such that $\tilde R_i =x$ for all $i \in C$ and (ii) there do not exist $C \in \mathcal C$ and $x \in \{a,b\}$ such that $\tilde R_i =x$ for all $i \in C$. In both cases, and consistent to Definition \ref{def:stpind}, to establish strategy-proofness we need only consider voters $i \in N$ such that $\tilde R_i \in \{a,b\}$.

In case (i), $\mathcal C^f(\tilde R_N)=x$. Suppose without loss of generality that $x=a$. Let $i \in N$ and suppose that $\tilde R_i =b$, meaning that $i \not \in C$. In this case, $\mathcal C^f(\tilde R_i',\tilde R_{-i})=a$ for any $\tilde R_i'\in \{a,b,0\}$. 

In case (ii), $\mathcal C^f(\tilde R_N)=f(\tilde R_N)$. Let $i \in N$ such that $\tilde R_i \in \{a,b\}$ and $\tilde R_i \neq f(\tilde R_N)$. Without loss of generality, suppose that $\tilde R_i=b$. Let $\tilde R_i' \in \{a,0\}$. 

Suppose first that $\tilde R_i'=a$. Then either (1) there exists $C \in \mathcal C$ such that $\tilde R_j = a$ for all $j \in C$, or (2) there is no such $C$, so that $\mathcal C^f(\tilde R_i',\tilde R_{-i})= f(\tilde R_i',\tilde R_{-i})$. In case (1), $\mathcal C^f(\tilde R_i',\tilde R_{-i}) =a$. Thus, $\mathcal C^f(\tilde R_i',\tilde R_{-i})$ is not preferred to $f(\tilde R_N)$. In case (2), $\mathcal C^f(\tilde R_i',\tilde R_{-i})=f(\tilde R_i',\tilde R_{-i})$. The strategy-proofness of $f$ ensures that $f(\tilde R_i',\tilde R_{-i})$ is not preferred to $f(\tilde R_N)$. Thus, in both cases $\mathcal C^f(\tilde R_i',\tilde R_{-i})$ is not preferred to $\mathcal C^f(\tilde R_N)$.

Suppose now that $\tilde R_i'=0$. Then $\mathcal C^f(\tilde R_i',\tilde R_{-i})=f(\tilde R_i',\tilde R_{-i})$. The strategy-proofness of $f$ ensures that $\mathcal C^f(\tilde R_i',\tilde R_{-i})=f(\tilde R_i',\tilde R_{-i})$ is not preferred to $\mathcal C^f(\tilde R_N)=f(\tilde R_N)$. 

We conclude that $\mathcal C^f$ is strategy-proof.

\bigskip

We now consider sequential unanimity rules. It is easy to show that they are neutral. To establish their strategy-proofness, consider a rule $\mathcal S^f$ with associated sequence S=$(S_1,...,S_K)$ and default $f$. We distinguish between two cases.

\medskip

\noindent {\bf Case 1:} There exist $k\in \{1,...,K\}$ and $x \in \{a,b\}$ such that (i) $\tilde R_i=x$ for all $i \in S_k$ and (ii) for all $l=1,..,k-1$ there exists $i_l \in S_l$ such that $\tilde R_{i_l}=x$. 

Suppose $k$ is the minimum such index and that, without loss of generality, $x=a$. Hence $\mathcal S^f(\tilde R_N)=a$. Consider $i\in N$ and suppose that $\tilde R_i =b$. This implies that $i\not \in S_k$. Let $\tilde R_i'\in \{a,0\}$. Then, repeating the reasoning in the proof of Proposition~\ref{prop:seq}, we obtain $\mathcal S^f(\tilde R_i',\tilde R_{-i})=a$. Thus, $i$ dot not prefer $\mathcal S^f(\tilde R_i',\tilde R_{-i})$ to $\mathcal S^f(\tilde R_N)$.

\medskip

\noindent {\bf Case 2:} There exist no $k\in \{1,...,K\}$ and $x \in \{a,b\}$ such that (i) $\tilde R_i=x$ for all $i \in S_k$ and (ii) for all $l=1,..,k-1$ there exists $i_l \in S_l$ such that $\tilde R_{i_l}=x$. 

Thus, $\mathcal{S}(\tilde R_N)=f(\tilde R_N)$. Let $i\in N$ and suppose without loss of generality that $\tilde R_i= b$. Let $\tilde R_i'\in \{a,0\}$. Suppose first that $\tilde R_i'=a$. There are two cases: either (I) there exists $k\in \{1,...,K\}$ such that (i) $\tilde R_i=a$ for all $i \in S_k$ and (ii) for all $l=1,..,k-1$ there exists $i_l \in S_l$ such that $\tilde R_{i_l}=a$, or (II) there exists no such $k \in \{1,...,K\}$. In case (I), $\mathcal S^f(\tilde R_i',\tilde R_{-i}) =a $. In case (II), $\mathcal S^f(\tilde R_i',\tilde R_{-i})=f(\tilde R_i',\tilde R_{-i})$ and the strategy-proofness of $f$ ensures that $i$ does not prefer $f(\tilde R_i',\tilde R_{-i})$ to $f(\tilde R_N)$. In either case, $i$ does not prefer $\mathcal S^f(\tilde R_i',\tilde R_{-i})$ to $\mathcal{S}^f(\tilde R_N)$.

Suppose now that $\tilde R_i'=0$. Then, $\mathcal S^f(\tilde R_i',\tilde R_{-i})=f(\tilde R_i',\tilde R_{-i})$ and the strategy-proofness of $f$ ensures that $i$ does not prefer $f(\tilde R_i',\tilde R_{-i})$ to $f(\tilde R_N)$. Thus, once again, $i$ does not prefer $\mathcal S^f(\tilde R_i',\tilde R_{-i})$ to $\mathcal{S}^f(\tilde R_N)$.

 We conclude that $\mathcal S^f$ is strategy-proof. \qed

\medskip

We continue by establishing analogues of Theorems~\ref{th:Alg1} and \ref{th:Alg2} for the full preference domain. In both cases, the proofs are straightforward adaptations of the arguments used previously. We include them for completeness.

\begin{theorem}
	Given an M-winning coalition set $\mathcal C$ and a neutral and strategy-proof rule $f$, consider the M-winning coalition rule $\mathcal C^f$. Furthermore, consider a sequential unanimity rule $\mathcal S^f$, whose associated sequence $\textrm S$ is an output of Algorithm 1 with $\mathcal C$ as input. For all $\tilde R_N \in \mathcal{\tilde R}^N$, we have $\mathcal C^f(\tilde R_N) = \mathcal S^f(\tilde R_N)$. 
\end{theorem}

\noindent {\bf Proof.} Consider the sequential unanimity rule $\mathcal S^f$ with associated sequence $\textrm S$, and let $\tilde R_N\in \mathcal{\tilde R}^N$. We distinguish between two cases:

	\noindent {\bf Case 1:}  For some $x \in \{a,b\}$, there exists $ C \in \mathcal C$ such that $\tilde R_i=x$ for all $i \in  C$. Then, by definition, $\mathcal C^f(\tilde R_N)=x$. 

	Consider the sequence $\textrm S=(S_1,...,S_K)$ produced by Algorithm 1 with $\mathcal C$ as input. By construction, there exists $k \in \{1,...,K\}$ such that $S_k \subset C$. We distinguish between two cases: (i) $C=S_k$ and (ii) $S_k \subsetneq C$.  In case (i), $S_l \in \mathcal C$ for all $l=1,...,k-1$. Hence, by property P2, $S_l \cap C \neq \emptyset$ for all $l=1,...,k-1$. In case (ii), for all $l=1,...,k-1$, either $S_l \in \mathcal C$, or $S_l \not \in \mathcal C$. If $S_l \in \mathcal C$, then, by property P2, $S_l \cap C \neq \emptyset$. If $S_l \not \in \mathcal C$, then condition (iii) in Step 2 of Algorithm 1 implies that $S_l \cap C \neq \emptyset$.
	
	Thus, in either case, for all $l<k$, there exists $i_l \in S_l \cap C$. Moreover, recall that $\tilde R_i=x$ for all $i \in C \supset S_k$. Hence, $\mathcal S^f(\tilde R_N)=x=\mathcal C^f(\tilde R_N)$. 
	
	\medskip
	
	\noindent {\bf Case 2:}  There does not exist any $C \in \mathcal C$ and $x \in \{a,b\}$, such that $\tilde R_i=x$ for all $i \in C$. In this case, $\mathcal C^f(\tilde R_N)=f(\tilde R_N)$. 
	
	Suppose that $\mathcal S^{f}(\tilde R_N)\neq f(\tilde R_N)$. Then, there exists $x \in \{a,b\}$ and  $k \in \{1,..,K\}$ such that (i) $\tilde R_i=x\neq f(\tilde R_N)$ for all $i\in S_k$ and (ii) for all $l=1,2,...,k-1$, there exists $i_l \in S_l$ such that $\tilde R_{i_l}=x$. By Theorem \ref{th:Alg2}, the set $\{i_1,i_2,.., i_{k-1}\} \cup S_k$, or a proper subset of it, will belong to the output M-winning coalition set which, by Corollary~\ref{coro:12}, equals $\mathcal C$. Thus, there exists $C \in \mathcal C$ such that $\tilde R_i=x$ for all $i \in C$, which contradicts the assumption of Case 2. Hence, $\mathcal S^{f}(\tilde R_N)=f(\tilde R_N)=\mathcal C^f(\tilde R_N)$. 
	
	\qed

\medskip

Finally, we prove the reverse result.

\begin{theorem}
	Given a sequence $\textrm S=(S_1,...,S_K)$ of subsets of $N$ satisfying $|S_K|=1$ and a neutral and strategy-proof rule $f$, consider the rule $\mathcal S^f$. Furthermore, consider the rule $\mathcal C^f$, whose associated M-winning coalition set $\mathcal C^S$ is the output of Algorithm 2 with $\textrm S$ as input. For all $\tilde R_N \in \mathcal{\tilde R}^N$, we have $\mathcal C^f(\tilde R_N) = \mathcal S^f(\tilde R_N)$. 
\end{theorem}

\noindent{\bf Proof.} Consider a profile $\tilde R_N$. By Lemma~\ref{lem:Mwin}, $\mathcal C^S$ is an M-winning coalition set. We distinguish between two cases:

		\noindent {\bf Case 1:} There exists $k\in \{1,...,K\}$ and $x \in \{a,b\}$ such that (i) for all $i \in S_k$, $\tilde R_i=x$ and (ii) for all $l=1,...,k-1$, there exists $i_l \in S_l$ such that $\tilde R_{i_l}=x$. 
	
	For simplicity, suppose that $k$ is the smallest index for which this statement is true and, without loss of generality that $x=a$. As a result, $\mathcal S^f(\tilde R_N)=a$. Following the same logic as the proof of Theorem~\ref{th:Alg2}, there exists $C\in \mathcal C^S$ such that $C \subset \{i_1,...,i_{k-1}\} \cup S_k$ . This implies $\mathcal C^f(\tilde R_N)=\mathcal S^f(\tilde R_N)=a$. 
	
	\medskip
	
		\noindent {\bf Case 2:} There exists no $k\in \{1,...,K\}$ and $x \in \{a,b\}$ such that (i) for all $i \in S_k$, $\tilde R_i=x$ and (ii) for all $l=1,...,k-1$, there exists $i_l \in S_l$ such that $\tilde R_{i_l}=x$.  
	
	Thus, $\mathcal S^f(\tilde R_N)= f(\tilde R_N)$. We make the following Claim.
	
	\paragraph{Claim 1.} {\em Assume the condition of case 2 holds.\footnote{That is, there exists no $k\in \{1,...,K\}$ and $x \in \{a,b\}$ such that: (i) for all $i \in S_k$, $\tilde R_i=x$ and (ii) for all $l=1,...,k-1$, there exists $i_l \in S_l$ such that $\tilde R_{i_l}=x$.} Then, there exists no $C\in \mathcal C^S$ and $x \in \{a,b\}$ such that $\tilde R_i=x$ for all $i \in C$.}
	
	\medskip
	\noindent{\bf Proof:} We prove the contrapositive statement. Suppose that there exists $C \in \mathcal C^S$ and $x \in \{a,b\}$ such that $\tilde R_i=x$ for all $i \in C$. Without loss of generality, suppose that $x=a$. By construction, there exists a subset $S_k$ and a path $p_k$ from $S_1$ to $S_k$ such that $C= \{p_k\} \cup S_k$. Thus, $\tilde R_i=a$ for all $i \in S_k$. Moreover, Definition \ref{def:path} implies that, for all $h=1,...,k-1$ such that $S_h \cap S_k = \emptyset$, there exists $i_h \in S_h \cap \{p_k\}$. Thus, for all $l=1,...,k-1$, there exists $i_l \in S_l$ such that either $i_l \in S_k$ or $i_l \in \{p_k\}$. Hence, for all $l=1,...,k-1$, there exists $i_l \in S_l$ such that $\tilde R_{i_l}=a$. Thus, there exists $k \in \{1,...K\}$ and $x \in \{a,b\}$ such that (i) for all $i \in S_k$, $\tilde R_i=a$ and (ii) for all $l=1,...,k-1$, there exists $i_l \in S_l$ such that $\tilde R_i=a$. This conclusion contradicts our assumption in Case 2 and thus ends the proof of Claim 1.

	\medskip
	
	\noindent Claim 1 implies that $\mathcal C^f(\tilde R_N)= \mathcal S^f(\tilde R_N)= f(\tilde R_N)$. Thus, the desired conclusion holds also in Case 2. \qed

\section{Conclusion}

In this paper we introduced a new family of rules, sequential unanimity rules, for voting between two alternatives. They are parametrized by a sequence of voter subsets such that the last element is a singleton. The first subset in the sequence in which voters are unanimous determines the outcome. Exploiting their intimate connection to M-winning coalition rules \cite{m83}, we proved that a subfamily of sequential unanimity rules are characterized by strategy-proofness and neutrality. We did so by developing algorithms that convert a given M-winning coalition rule into an equivalent essential sequential unanimity rule and vice versa. Since strong simple games are formally identical to M-winning coalition rules, these results are potentially of interest to the literature on strong simple games as well. Finally, we extended the reach of our analysis to cover the full preference domain which allows for indifferences. 


\section*{Appendix}

\subsection*{A1: Computation of sequences $\textrm S^2$ and  $\textrm S^3$ of Section 3.1.2}

In this section, we demonstrate how Algorithm 1 produces sequences $\textrm S^2$ and $\textrm S^3$ cited in the main text.

\begin{figure}[H]
	\centering
	\includegraphics[height=.2\textheight, width=.75\textwidth]{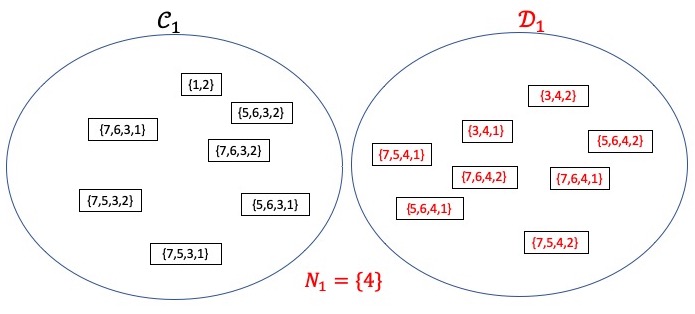}
	\caption{{\small Iteration 1: Set backstop voter $i=4$ and remove all elements of $\mathcal C$ containing it. The remaining M-winning coalition set is $\mathcal C_1$ and the discarded M-winning coalition set is $\mathcal D_1$.}}
	\label{fig:Alg1_2_2}
\end{figure}

We first address sequence  $\textrm S^2$. We begin by selecting voter 4 as the backstop, so that $N_1=\{4\}$. Subsequently, we update the M-winning coalition set by removing all its elements containing voter 4. Figure \ref{fig:Alg1_2_2} illustrates.

\begin{figure}[H]
	\centering
	\includegraphics[height=.25\textheight, width=.75\textwidth]{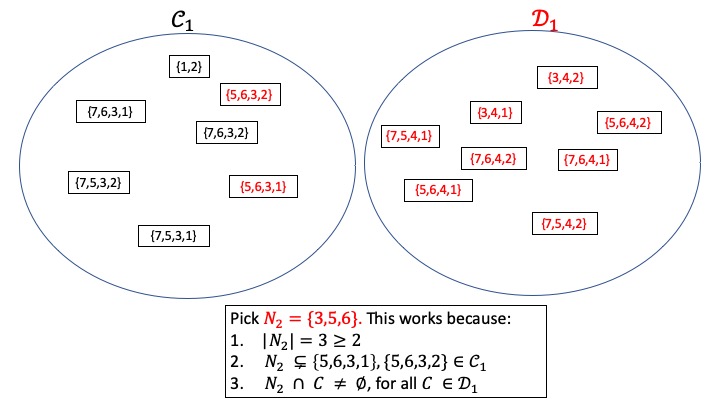}
	\caption{{\small Iteration 2: Selecting subset $N_2$.}}
	\label{fig:Alg1_2_3}
\end{figure}

We continue by searching for a subset of voters $N_2$ that satisfies the three criteria laid out in the previous subsection. There are various choices we could make at this point and we choose $N_2=\{3,5,6\}$. Figure \ref{fig:Alg1_2_3} highlights the elements of $\mathcal C_1$ that are proper supersets of $N_2$ and demonstrates why the subset $\{3,5,6\}$ is a is a valid choice. For conciseness, we omit explicitly displaying the corresponding updates of the remaining M-winning coalition set $\mathcal C_1$ and the discarded M-winning coalition set $\mathcal D_1$, to $\mathcal C_2$ and $\mathcal D_2$ respectively. 


\begin{figure}[H]
\centering
\includegraphics[height=.25\textheight, width=.75\textwidth]{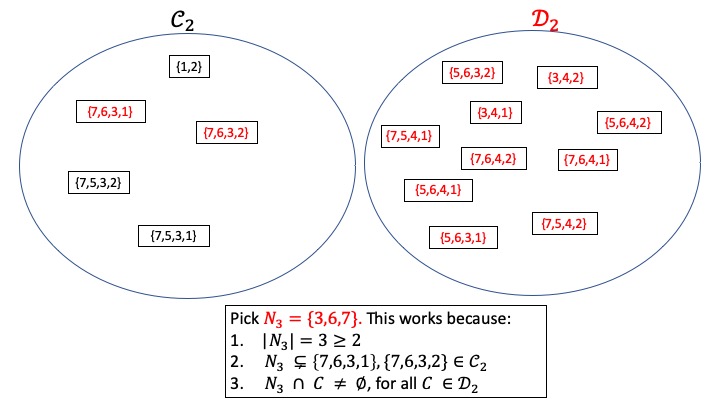}
\caption{{\small  Iteration 3: Selecting subset $N_3$.}}
\label{fig:Alg1_2_5}
\end{figure}

We continue by choosing $N_3=\{3,6,7\}$. Figure \ref{fig:Alg1_2_5} illustrates why this is a valid selection. 



\begin{figure}[H]
\centering
\includegraphics[height=.25\textheight, width=.75\textwidth]{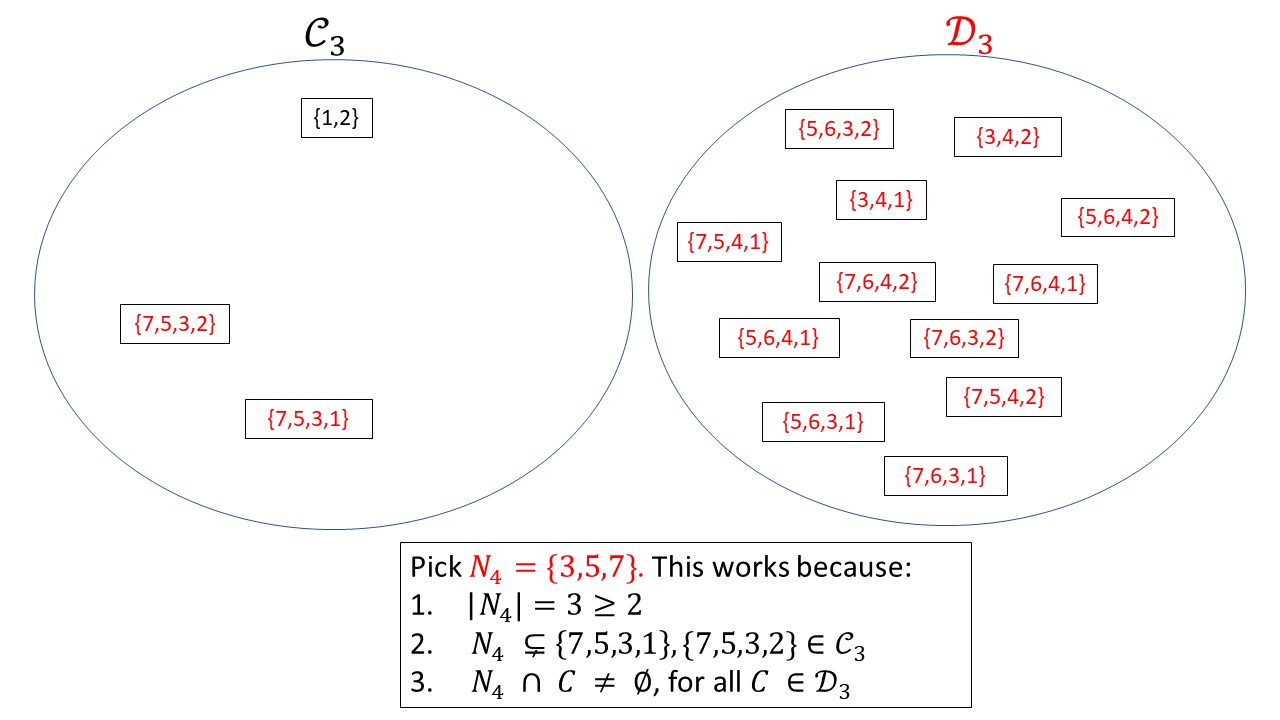}
\caption{{\small  Iteration 4: Selecting subset $N_4$.}}
\label{fig:Alg1_2_6}
\end{figure}

We continue by choosing $N_4=\{3,5,7\}$. Once again, Figure \ref{fig:Alg1_2_6} illustrates why this is a valid selection.  

%

After iteration 4, there is a single element left in the remaining M-winning coalition set $\mathcal C_4$. Moreover, there is no way to satisfy all three criteria: in particular, criteria (i) and (ii) are mutually exclusive. Thus, Algorithm 1 terminates. The output sequence that it produces is the following:$$\textrm S^2=(\{1,2\}, N_4, N_3, N_2, N_1)= ( \{1,2\}, \{3,5,7\}, \{3,6,7\}, \{3,5,6\},\{4\}).$$

\begin{figure}[H]
\centering
\includegraphics[height=.2\textheight, width=.75\textwidth]{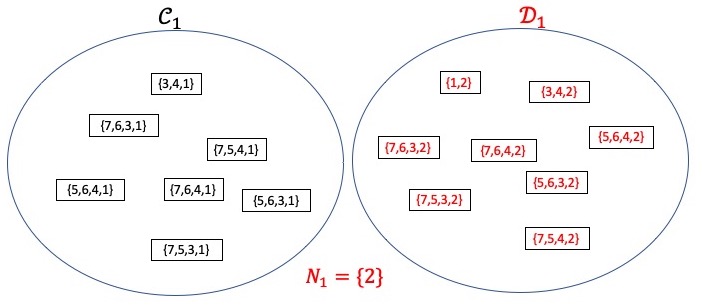}
\caption{Iteration 2: After choosing voter 2 as the backstop, no subset $N_2$ satisfies the three criteria.}
\label{fig:Alg1_3_2}
\end{figure}

Finally, we show what happens when the algorithm terminates with multiple remaining subsets. To make this point as stark as possible, we select voter 2 as the backstop voter, so that $N_1=\{2\}$. Subsequently, we remove all elements of $\mathcal C$ that contain voter 2, and obtain $\mathcal C_1$ and $\mathcal D_1$ as depicted in Figure \ref{fig:Alg1_3_2}.

At this point there is no subset of $N$ that satisfies criteria (i)-(ii)-(iii). For example, subset $\{5,6\}$ satisfies (i) and (ii) but fails (iii): we have $\{1,2\} \in \mathcal D_1$ and $\{5,6\} \cap \{1,2\} = \emptyset$. Or, similarly, subset $\{1,3,6\}$ satisfies (i) and (ii) but fails (iii): we have $\{7,5,4,2\} \in \mathcal D_1$ and $\{1,3,6\} \cap \{7,5,4,2\} = \emptyset$. Thus, the algorithm terminates, with seven elements of $\mathcal C_1$ still remaining. These seven subsets can be ordered in an arbitrary fashion and placed in the first seven spots in the output sequence. Choosing one such ordering, we obtain the following sequence:$$\textrm S^3=(\{7,5,3,1\}, \{5,6,3,1\}, \{7,5,4,1\}, \{5,6,4,1\},\{7,6,3,1\},\{3,4,1\}, \{7,6,4,1\}, \{2\}).$$


\footnotesize{
}

\end{document}